\documentclass[twocolumn,showpacs,preprintnumbers,amsmath,amssymb]{revtex4}

\usepackage{graphicx}
\usepackage{dcolumn}
\usepackage{bm}

\def\bea{\begin{eqnarray}}
\def\eea{\end{eqnarray}}

\begin{document} 

\preprint{Version 1.7}

\title{Centrality evolution of $p_t$ and $y_t$ spectra from Au-Au collisions at $\sqrt{s_{NN}} = 200$ GeV}

\author{Thomas A. Trainor}
\address{CENPA 354290, University of Washington, Seattle, WA 98195}


\date{\today}

\begin{abstract}
A two-component analysis of spectra to $p_t = 12$ GeV/c for identified pions and protons from 200 GeV Au-Au collisions is presented. The method is similar to an analysis of  the $n_{ch}$ dependence of $p_t$ spectra from p-p collisions at 200 GeV, but applied to Au-Au centrality dependence. The soft-component reference is a L\'evy distribution on transverse mass $m_t$. The hard-component reference is a Gaussian on $y_t$ with exponential ($p_t$ power-law) tail. Deviations of data from the reference are described by hard-component ratio $r_{AA}$ which generalizes nuclear modification factor $R_{AA}$. The analysis suggests that centrality evolution of pion and proton spectra is dominated by changes in parton fragmentation. The structure of $r_{AA}$ suggests that parton energy loss produces a negative boost $\Delta y_t$ of a large fraction (but not all) of the minimum-bias fragment distribution, and that lower-energy partons suffer relatively less energy loss, possibly due to color screening. The analysis also suggests that the anomalous $p/\pi$ ratio may be due to differences in the parton energy-loss process experienced by the two hadron species. This analysis provides no evidence for radial flow.
\end{abstract}


\maketitle

 \section{Introduction}

$p_t$ spectrum analysis of relativistic heavy ion collisions typically invokes the assumption that a thermally-equilibrated flowing bulk medium is produced. The hydrodynamic (hydro) model plays a central role in the description~\cite{teaney,pasi,kolb}. Hard-scattered partons are proposed as probes of the medium~\cite{tomo}. A thermalized system is said to expand and cool until chemical and kinetic decoupling. $p_t$ spectra for identified hadrons are used to derive thermodynamic and chemical properties of the medium. Blast-wave fits and mass dependence are used to infer radial flow~\cite{expand,kolb2}. Temperatures and chemical potentials are inferred from statistical-model spectrum fits and particle ratios~\cite{ratio}. Unexpected spectrum properties motivate supplemental recombination or coalescence models~\cite{hwarec,friesrec,grecrec}. From the combined spectrum information evidence for a QCD phase transition and formation of a quark-gluon plasma is sought.

Observationally, $p_t$ spectra for identified hadrons from heavy ion collisions at RHIC extending to $p_t = 12$ GeV/c as in this analysis  may vary over 10 orders of magnitude, whereas relative data accuracy may be 5-10\% over much of the spectrum. It is thus difficult to represent visually the full information content of carefully measured spectra. According to conventional assumptions $p_t$ is separated into a thermal/hydro region $p_t < 2$ GeV/c, a coalescence/recombination (ReCo) region $2 < p_t < 6$ GeV/c and a perturbative QCD (pQCD) region $p_t > 6$ GeV/c. Each interval receives a separate analysis strategy and theoretical treatment. One can question whether all physical information in the spectra is being used to test theory. In this paper I show that differential spectrum analysis is essential to exploit the data fully. 

Analysis of the multiplicity dependence of $p_t$ spectra from NSD p-p collisions revealed that the {\em two-component model} of $p_t$ spectra provides an accurate and complete description of p-p spectrum data for unidentified hadrons~\cite{ppprd}. The soft-component shape is independent of multiplicity. The hard component {\em shape} is independent of multiplicity, but its {\em amplitude relative to the soft component} is proportional to multiplicity. The p-p hard component is approximately a Gaussian on transverse rapidity $y_t$, and its form is related both to parton momentum spectra and to parton fragmentation functions in $e^+$-$e^-$ collisions~\cite{lepmini}. 

The hard component of p-p spectra can be interpreted as a minimum-bias {\em parton fragment distribution} (FD)---conditional fragmentation functions folded with the unbiased parton spectrum. In that interpretation a substantial fraction of the full hadron spectrum down to $p_t \sim 0.3$ GeV/c is part of the parton fragment distribution, with implications for A-A $p_t$ spectra. Parton fragmentation then competes with hydro, thermalization and recombination to account for the bulk of A-A particle production. To sort out competing mechanisms we must understand A-A $p_t$ spectra to the {statistical limits of the data}, with as little physical model dependence as possible. 

In this paper I apply the two-component spectrum model to pion and proton spectra from five centrality classes of Au-Au collisions at $\sqrt{s_{NN}} = 200$ GeV~\cite{hiptspec}. Spectra are compared to a model function, extended from the two-component analysis of p-p collisions, which describes all structure in the data. 

The paper is organized as follows. I introduce the data used and discuss some conventional aspects of spectrum analysis. I describe the two-component spectrum model as applied to p-p and Au-Au collisions. I discuss differential spectrum analysis, including conventional nuclear modification factor $R_{AA}$ and its generalization in this analysis to hard-component ratio $r_{AA}$. I also introduce a simple parton energy-loss model including attenuation and negative boost of the minimum-bias fragment distribution.

I apply the two-component model to pion and proton spectra, obtaining soft- and hard-component model functions for each species and five centralities. I then obtain hard-component ratios $r_{AA}$ for the data which generalize $R_{AA}$ and $R_{CP}$ results. I model the data $r_{AA}$ trends with simple functional forms which suggest that parton energy loss produces a negative boost of part of the minimum-bias fragment distribution (the reference hard component). I compare those results to theory and observe that the data suggests the presence of color screening for lower-energy partons, explaining the large abundance of minijets observed with correlation analysis in central Au-Au collisions.

Finally, I demonstrate that the anomalous $p/\pi$ ratio may be an aspect of parton energy loss differently manifested for pions and protons, compare integrals of the spectrum models to the Kharzeev and Nardi (K-N) two-component model ~\cite{nardi}, and obtain centrality trends of $\langle p_t \rangle$ which are compared to published values.

\section{Conventional $p_t$ Spectrum Analysis}


The invariant single-particle density on 3D momentum space for identified hadrons  at $\eta = 0$ is
\bea  \label{eq1}
\rho(x_t) &\equiv&1/2\pi\,\, 1/x_t\,\, d^2n/dx_t\, d\eta  
\eea
averaged over $2\pi$ azimuth and one unit of pseudorapidity, with transverse variable $x_t = p_t$, $m_t$ or $y_t$ for momentum, mass or rapidity respectively. The Jacobian for transformation from $p_t$ to $y_t$ is $p_t\, m_t / y_t$, that for $p_t \rightarrow m_t$ is unity. Although the data analyzed here are for identified hadrons I use pseudorapidity to provide convenient comparisons with unidentified-hadron spectra.


Ratio techniques have been introduced to study ``jet quenching''~\cite{quench}. The conventional ratio measures (nuclear modification factors) are
\bea
R_{AA} &=& \frac{1}{n_{bin}}\cdot  \frac{ \rho_{AA}}{\rho_{pp}} \\ \nonumber
R_{CP} &=& \frac{n_{binP}}{n_{binC}}\cdot  \frac{\rho_C}{\rho_P},
\eea
the latter applied to more peripheral (P) collisions relative to more central (C) collisions when p-p reference spectra are unavailable for comparison.
The ratio limits for $p_t$ small are $1/\nu$ or $\nu_P / \nu_C$, where mean participant pathlength $\nu \equiv 2n_{binary} / n_{participant}$~\cite{centmeth}. Some features of the spectra at smaller $p_t$ are suppressed by those definitions. One purpose of this analysis is to restore access to those features.

\subsection{200 GeV Au-Au PID $p_t$ spectra}

The particle-identified (PID) data used in this analysis are large-statistics identified pion and proton $p_t$ spectra extending to $p_t\sim$12 GeV/c for five Au-Au centralities~\cite{hiptspec}.  Spectra were obtained from 15M 0-12\% central Au-Au collision events and 14M minimum-bias events at $\sqrt{s_{NN}} = 200$ GeV. The centralities are 0-12\%, 10-20\%, 20-40\%, 40-60\% and 60-80\%, corresponding to mean participant pathlength $\nu = 5.50, 4.87, 3.93, 2.83$ and $1.93$ respectively~\cite{centmeth}. Spectra are in the form of invariant particle density $\rho(p_t) = \rho(x_t)$ in Eq.~(\ref{eq1}) with $x_t \rightarrow p_t$. 

\subsection{Assumptions and interpretations}

The analysis in~\cite{hiptspec} represents a conventional $p_t$ spectrum analysis. By hypothesis spectra are divided into three $p_t$ intervals: 1) Particle production in $p_t < 2$ GeV is said to be dominated by soft QCD processes and described by hydro models assuming local thermalization and collective flow~\cite{teaney,pasi,kolb}. 2) In $p_t \in [2,6]$ GeV/c differing meson and baryon {\em suppression} is interpreted in terms of hadronization {\em via} ``coalescence of constituent quarks from a collective partonic system'' ~\cite{hwarec,friesrec,grecrec}. 3) For $p_t > 6$ GeV/c  particle production is assumed to reflect pQCD parton scattering, parton energy loss and fragmentation. 

Two $p_t$ intervals are relevant to this analysis.

1) The intermediate-$p_t$ region:
The stated issue is {\em relative suppression} of protons and pions. Protons are said to be ``less suppressed'' than pions, but significant suppression is still observed, in contrast to d-Au collisions where significant enhancement (Cronin effect) is observed~\cite{cronin}. Differences between proton and pion suppression (anomalous p/$\pi$ ratio) motivate the quark coalescence or recombination model, predicated on the prior existence of a thermalized collective partonic medium (QGP)~\cite{hwarec,friesrec,grecrec}.

2) The large-$p_t$ region:
Hadron yields are strongly suppressed in this $p_t$ interval, and pQCD theory attempts to predict the suppression in terms of parton energy loss. A major issue is whether suppression of different hadron species can reveal color or flavor dependence of parton energy loss~\cite{dok,elosstheor2}. Pion and proton suppression are similar, and the p/$\pi$ ratio returns to that for d-Au, suggesting that collectivity and coalescence are not important in this interval.

\section{Two-component spectrum model}

The two-component spectrum model is the extension of a model of hadron production in nuclear collisions~\cite{2comp,nardi}. To describe $p_t$ spectra hadrons are assumed to emerge from two processes, soft and hard, distinguishable by parametric variations (e.g., centrality $\nu$ or $n_{ch}$). Each process has a characteristic functional form on $p_t$ or $y_t$ which is approximately independent of collision parameters, but the relative amplitude of the two processes varies smoothly with a control parameter, providing the basis for separation. In the present paper the analysis is treated as a model-independent Taylor expansion of  density $\rho(y_t;\nu)$ on $\nu$, with no physical model assumptions. The terms of the series can then be identified with the conventional physics-based two-component model.

\subsection{Transverse rapidity $y_t$}  \label{tranrap}

$p_t$ spectra provide information about parton fragmentation, but the fragmentation process is more {\em visually} accessible on a logarithmic momentum variable~\cite{lepmini}. The dominant hadron (pion) is ultrarelativistic at 1 GeV/c, and a relativistic kinematic variable is a reasonable alternative. Transverse rapidity in a longitudinally comoving frame near midrapidity $y_z = 0$ is defined by
\bea
y_t = \ln([m_t + p_t] / m_0),
\eea
with $m_0$ a hadron mass and $m_t^2 = p_t^2 + m_0^2$. $\rho(y_t)$ is defined by $x_t \rightarrow y_t$ in Eq.~(\ref{eq1}); transformation of data densities $\rho(p_t)$ to $\rho(y_t)$ employs the Jacobian below Eq.~(\ref{eq1}). For pions (corresponding to all $y_t$ values in this paper) $p_t = 0.16,\, 0.5,\, 1,\, 2,\, 4,\, 6,\, 10$ GeV/c transforms to $y_t = 1,\, 2,\, 2.66,\, 3.33,\, 4,\, 4.5,\, 5$ respectively.

The initial version of this analysis defined separate $y_t$s for pions and protons according to their masses ($m_0$) to accommodate a radial flow component according to the Cooper-Frye formalism~\cite{cooper}, with a common positive boost $\Delta y_t$ (blue shift) of nearly-thermal spectra on transverse rapidity. However, there is {no evidence for hydro structure} in the spectra, and the hard component for protons was found to be approximately the same on $p_t$ as that for pions. Thus, $y_t$ was subsequently defined strictly as a logarithmic transformation from $p_t$ using the pion mass for both pions {\em and} protons. The transformation increases visual access to spectrum structure but has no other physical significance. For thermal production from a boosted source the proper $y_t$ for each hadron would be required, and spectra should be considered in general with both $y_t$ definitions in mind. 

\subsection{p-p two-component model}   \label{ppmodel}

In the two-component spectrum model applied to p-p collisions at 200 GeV the soft-component density ($\sim$ longitudinal {\em nucleon} fragmentation) is simply proportional to the integrated event multiplicity $n_{ch}$ within $\eta \in [-0.5,0.5]$ ($ n_{ch} \sim d n/d\eta \equiv  \rho$, the 1D density on $\eta$), and the hard-component density ($\sim$ transverse {\em parton} fragmentation) is proportional to the multiplicity squared~\cite{ppprd}. $y_t$ spectra from p-p collisions are accurately described by $\rho_{pp}(y_t) = \rho_s(n_{ch}) S_0(y_t) + \rho_h(n_{ch}) H_0(y_t)$, with $\rho_h/\rho_s \equiv x \sim 0.005\, n_{ch}$ and $\rho_h + \rho_s = \rho$ ($\rho_s$ and $\rho_h$ replace $n_s$ and $n_h$ from~\cite{ppprd}). $S_0$ is a L\'evy distribution on $m_t$ and $H_0$ is a Gaussian on $y_t$, each normalized to unity. 
In the notation of the present analysis
\bea \label{pp2comp}
\rho_{pp}(y_t;x) &=& \rho\, \{S_0(m_t[y_t]) + x\, H_0(y_t)\}/(1 + x).
\eea
Event multiplicity $ n_{ch}$  serves in effect as a hard trigger. $\bar \rho = \bar n_{ch} \sim 2.5$ is the $p_t$-integrated 1D NSD hadron density on $\eta$ averaged over $2\pi$ azimuth, and $\eta \in [-0.5,0.5]$ at 200 GeV. Because the mean minijet multiplicity and $dn/d\eta$ for NSD p-p collisions are both $\sim 2.5$,  $\bar x$ is approximately the probability to observe one hard parton scatter in $\eta \in [-0.5,0.5]$ in one NSD p-p collision.  

Soft-component model function $S_0$ is a unit-normal  L\'evy distribution~\cite{wilk}
\bea
S_0(m_t;n_s,T) &=&   A_s / \left[ 1 + \frac{m_t - m_0}{n_sT} \right]^{n_s},
\eea
an apparently minor modification ($m_t$ vs $p_t$) of the ``power-law'' distribution $A/(1+p_t/p_0)^n$~\cite{ua1}, but representing ``soft'' rather than ``hard'' processes (longitudinal rather than transverse fragmentation)~\cite{ppprd}. 
The density on $m_t$ is defined with the proper pion or proton mass, but the transformation to $S_0(y_t)$ is then made with the same pion $m_t$ and $y_t$ used for both protons and pions.

Hard-component model function $H_0$ in~\cite{ppprd} is a unit-normal Gaussian on $y_t$
\bea
H_0(y_t;\bar y_t, \sigma_{y_t}) = A_h \exp\left\{\frac{-(y_t - \bar y_t)^2}{2\sigma^2_{y_t}}\right\}.
\eea
As with $S_0$ the form is independent of control parameter $n_{ch}$. In this Au-Au analysis the common $y_t$ with pion mass is used for protons and pions, but the Gaussian parameters are found to be somewhat different for the two hadron species. 

For this analysis $\rho_0 = d^2n/d\eta\, d\phi = 2.5/2\pi$ is the average $p_t$-integrated 2D hadron density on ($\eta,\phi$) at mid-rapidity. The $p_t$-integrated soft and hard components are $\bar \rho_s = \rho_0 / (1+\bar x)$ and  $\bar \rho_h = \rho_0\,  \bar x/ (1+\bar x)$, with $\bar x = 0.012$ for NSD p-p collisions at 200 GeV. I then define $S_{pp}(y_t) = \bar \rho_s\, S_0(y_t)$, $H_{pp}(y_t) = \bar \rho_h\, H_0(y_t)$, and the multiplicity-dependent {\em per-particle} 3D density is
\bea
\bar \rho_s/\rho_s\cdot \rho_{pp}(y_t;n_{ch}) = S_{pp}(y_t)  + n_{ch}/\bar n_{ch} \cdot H_{pp}(y_t).
\eea
That expression for p-p collisions is formally equivalent to Eq.~(\ref{aa2comp}) for A-A collisions.

\subsection{$H_0$ and the power-law tail}

The spectra used for the two-component analysis in~\cite{ppprd} extended only to $p_t \sim$ 6 GeV/c. The principal result was isolation of the hard component of the p-p $p_t$ spectrum as a Gaussian on $y_t$ describing parton fragment distributions~\cite{lepmini}. In the present analysis the spectra extend to $p_t = 12$ GeV/c, and the form of $H_0$ must be modified. The QCD power-law trend, implemented previously by a monolithic ``power-law'' function~\cite{ua1}, is implemented in this analysis within modified hard-component model function $H_0$. The method to generate $H_0$ as a Gaussian on $y_t$ with exponential tail consistent with pQCD power law $p_t^{-n_h}$ is described in App.~\ref{power}. 

The pQCD exponent $n_h$ associated with the underlying parton $p_t$ spectrum should be clearly distinguished from the model parameter $n_s$ associated with the soft-component L\'evy distribution. In Fig.~2 (right panel) of~\cite{ppprd} the variation of ``power-law'' fitting parameter $n$ with p-p event multiplicity ranges from $n_s \in [14,20]$ for small multiplicities to $n_h \in [7,8]$ for large multiplicities: the hard component increasingly dominates the spectrum with increasing multiplicity ({\em cf.} Fig.~10, left panel of~\cite{ppprd}). The conventional ``power-law'' function models two physical mechanisms with one functional form. Its exponent $n$ is oversubscribed, representing one process at smaller $p_t$ or $n_{ch}$ and another process at larger $p_t$ or $n_{ch}$. In the present analysis we observe $n_h \sim 7.5$, consistent with QCD expectations. Because of the Jacobian between $p_t$ and $y_t$ the exponential parameter in $H_{0}(y_t;\bar y_t,\sigma_{y_t},n_{y_t})$ is $n_{y_t}\sim 5.5$, as described in App.~\ref{power}.

\subsection{A-A two-component model}

The two-component model of hadron production (spectrum integrals) in A-A collisions assumes that the soft component is proportional to the participant pair number (linear superposition of N-N collisions), and the hard component is proportional to the number of N-N binary collisions (parton scattering)~\cite{nardi}. Any deviations from the model are specific to A-A collisions and may reveal properties of an A-A medium. In terms of mean participant path length $\nu = 2n_{bin} / n_{part}$~\cite{centmeth} the $p_t$-integrated Au-Au hadron density on $\eta$ is 
\bea \label{kn2comp}
\frac{2}{n_{part}}\frac{dn}{d\eta}_{AA}  &=& \rho_{s} +  \nu\, \rho_{h}, \\ \nonumber
&=& \rho_0\, [1 + x_\text{KN}\, (\nu - 1)],
\eea
with $   x_\text{KN} =\rho_{h}/ \rho_{0}$ and $x_{KN} \sim 0.08$ for 130 GeV and 0.10 for 200 GeV Au-Au collisions~\cite{nardi}. From the p-p two-component analysis we expect $x_{KN} \rightarrow x = 0.012$, which seems to conflict with~\cite{nardi}. I resolve the apparent contradiction in Sec.~\ref{partprod}.

In this analysis I extend the two-component particle-production model by analogy with Eq.~(\ref{kn2comp}) to describe Au-Au single-particle spectra on $p_t$ and $y_t$. As with the p-p analysis the model is essentially the first terms of a Taylor expansion on A-A centrality parameter $\nu$
\bea  \label{aa2comp}
\frac{2}{n_{part}} \rho_{AA}(y_t;\nu) &=& S_{NN}(y_t) +  \nu\, H_{AA}(y_t;\nu) \\ \nonumber
&=&  S_{NN}(y_t) +  \nu\,r_{AA}(y_t;\nu) \,H_{NN}(y_t).
\eea
Ideally, the A-A reference functions would be p-p limiting cases $S_{pp}$ and $H_{pp}$. But there are small differences depending on what hadron species or combinations are used to define $X_{pp}$ and $X_{NN}$, and possible isospin effects. A-A limiting cases are therefore specified in terms of N-N collisions. 

With those exceptions the shape of $S_{NN} \sim S_{pp}$ appears not to change significantly with A-A centrality. For peripheral collisions we expect $H_{AA} \rightarrow  H_{NN}$. For more central collisions $H_{AA}$ is strongly modified, revealing parton energy-loss systematics which are absorbed into factor $r_{AA}$, the ratio of $H_{AA}$ to $H_{NN}$.

\section{Differential spectrum analysis}

Conventional methods of differential spectrum analysis emphasizing the hard component at large $p_t$ are compared  with alternative methods which reveal structure over the entire $p_t/y_t$ acceptance.

\subsection{Conventional ratio measures}

Conventional spectrum ratios are expressed in two-component form as
\bea
R_{AA} &=& \frac{1}{\nu}\cdot  \frac{S_{NN} +\nu\, H_{AA}}{S_{pp} + H_{pp}} \\ \nonumber
R_{CP} &=& \frac{\nu_P}{\nu_C}\cdot  \frac{S_C + \nu_C\, H_{C}}{S_P + \nu_P \, H_{P}}.
\eea
Effects of the medium on hard processes are measured at large $p_t$ by the deviation of $R_{AA}$ from unity according to~\cite{star-raa}. The alternative form
\bea
\nu\, R_{AA} &=&   \frac{S_{NN} + \nu \,H_{AA}}{S_{pp} + H_{pp}},
\eea
should go asymptotically to 1 at small $p_t$ or $y_t$ and to  $\nu$ at large $p_t$ or $y_t$. I compare data in both forms to their two-component references. Since soft components $S_*$ are approximately equal the limits of $R_*$ for $p_t$ small, where $S_*$ dominates, are $1/\nu$ or $\nu_P / \nu_C$. However, important hard structure may persist at small $p_t$ which is strongly suppressed by the $R_*$ definitions. 

\subsection{Isolating the hard component}

To extract the hard component from data I define the difference
\bea  \label{hard}
 H_{AA} &\equiv& \frac{1}{\nu} \left\{\frac{2}{n_{part}} \rho_{AA} - S_{NN}\right\} .
\eea
As noted, I adopt $S_{NN} \sim S_{pp}$ from~\cite{ppprd} but optimize the model parameters for Au-Au data and the specific hadron species, so that $2/n_{part} \cdot \rho_{AA}(y_t) \rightarrow S_{NN}$ as $\nu \rightarrow 0$. The difference reveals hard component $H_{AA}$ for all $p_t$ or $y_t$ within the acceptance---{all parton fragments}, not just the pQCD region of the spectrum. In turn, I expect $H_{AA} \rightarrow H_{NN}$ as $\nu \rightarrow 1$. The limiting cases provide consistency checks on the analysis. An analogous procedure with the limit $\hat n_{ch} \rightarrow 0$ was applied in~\cite{ppprd} to isolate $S_{pp}$ and $H_{pp}$.

\subsection{Parton energy-loss model}

A simple two-parameter model of the effect of parton energy loss on the hard component is
\bea  \label{elossmodel}
H_{AA}(y_t;\nu) &=& A(\nu)\, H_{NN}[y_t + \Delta y_t(\nu)],
\eea
with $A < 1$ representing uniform attenuation and $\Delta y_t < 0$ representing a negative boost or {\em red shift} on $y_t$ (uniform {\em fractional} momentum reduction). I then define logarithmic difference
\bea  \label{elossmodel2}
 \Delta \log\{H_{AA}(y_t;\nu)\} = \log \left\{\frac{H_{AA}(y_t;\nu)}{H_{NN}(y_t)}\right\} 
\eea
as a generalization of ratio $R_{AA}$ which in contrast reveals details of fragmentation systematics over all $p_t$ or $y_t$. Differential analysis involves comparison of data to reference Eq.~(\ref{elossmodel}). Access to fragment distribution ratios over the entire $y_t$ acceptance, as illustrated in Sec.~\ref{ratios}, should improve understanding of parton energy loss and fragmentation.

\subsection{Plotting formats}

Because of their great numerical range and strong curvature $p_t$ spectra extending to large $p_t$ values don't provide sufficient visual resolution to reveal all significant physical information. Dot sets displaced by powers of ten strongly suppress differential structure. More information is revealed by improved plotting strategies and differential comparisons with a two-component reference. 

Transverse rapidity $y_t$ has three advantages over $p_t$ for a differential analysis: 1) Parton fragment distributions have a simple, nearly-symmetric form on $y_t$~\cite{lepmini}. 2) The Jacobian for $p_t \rightarrow y_t$ removes a factor $p_t^2$ at larger $y_t$, reducing the vertical range in this analysis by nearly three orders of magnitude and improving sensitivity to differential structure. 3) QCD power law $1/p_t^{n_h}$ at larger $p_t$ transforms to a linear trend in a semilog plot on $y_t$. 

Spectra are plotted as curves (line segments connecting data points) with common factor $n_{part}/2$ removed to form per-participant-pair densities. For participant scaling such curves would coincide for all centralities. Variations unique to heavy ion collisions appear as differences between the curves. Only plots with $H_{AA}$ isolated are sufficiently differential to reveal detailed structure. In those plots each centrality is identified by a specific line style. In other plots only the general relation of data (solid curves) to the two-component reference is shown. 

The dotted model curves in each plotting space provide a reference grid describing  the two-component model. Plotting scales are coordinated between equivalent plots for pions and protons to aid comparison. This analysis reveals simple systematic deviations of data from the model. Correspondence of spectrum variations across the full $y_t$ acceptance are made apparent for the first time.

\section{Two-component analysis -- Pions}

$p_t$ spectra for identified charged pions from five centralities of 200 GeV Au-Au and hadron spectra from NSD p-p collisions are compared in several plotting formats, and the hard components are isolated. Au-Au spectra were renormalized from the published data as described in App.~\ref{renorm}. The spectra from $\pi^+$ and $\pi^-$ were combined, since $\pm$ ratios show no structure on $p_t$.

\subsection{Pion spectra and soft component}

\begin{figure}[h]
  \includegraphics[width=3.3in,height=3.3in]{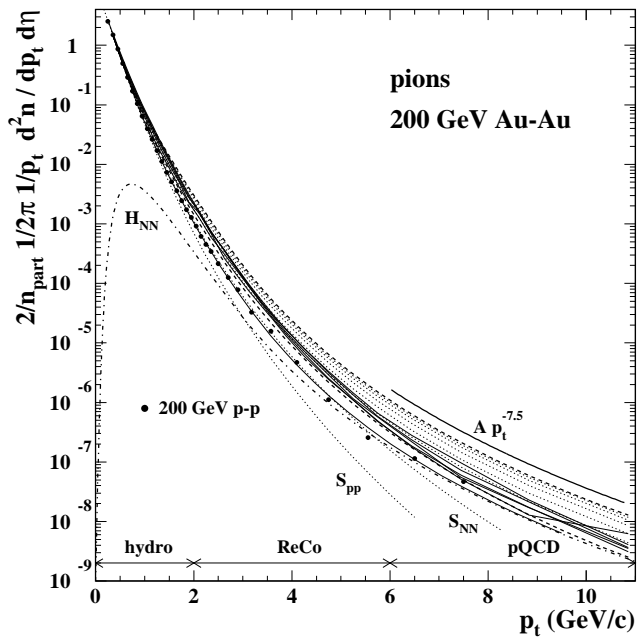}
\caption{\label{pipt} Pion $p_t$ spectra for five Au-Au centralities (solid curves). The dotted curves are corresponding two-component model functions, bounded by dashed curves for N-N and $b = 0$ Au-Au collisions. The dash-dot curve is the N-N hard-component reference function. The dotted curves labeled $S_{pp}$ and $S_{NN}$ are soft-component models for p-p and Au-Au collisions. The points are data from NSD p-p collisions (the solid curve through points is the two-component p-p model). A power-law trend $p_t^{-7.5}$ is shown at the right side of the plot. 
 } 
 \end{figure}

In Fig.~\ref{pipt} $p_t$ spectra for pions from five Au-Au centralities are plotted in the conventional format (five solid curves)~\cite{hiptspec}. The solid dots are the $\hat n_{ch} = 1$ spectrum from 3M NSD p-p collisions~\cite{ppprd}. N-N hard component $H_{NN}$ (dash-dot curve) is defined in Sec.~\ref{pionhard}, including a power-law tail as described in App.~\ref{power}. $S_{pp}$ and $H_{pp}$ (the latter not shown) from~\cite{ppprd} are redefined based on this analysis and combined to form the solid curve passing through the p-p points.

Soft component $S_{NN}$ (labeled dotted curve) has the same L\'evy form as for p-p collisions. The Au-Au pion parameters are $A_s = 20.2\pm0.1,\, T = 0.1445\pm0.001~\text{GeV, and}~ n_s = 12.0\pm0.2$. Exponent $n_s = 14.5$ with $A_s = 21.1$ is used for the p-p hadron spectrum in this analysis. The p-p exponent obtained in~\cite{ppprd} was $n_s = 12.8$. However, that analysis described hadrons using a Gaussian hard component with no power-law tail. In the present analysis with power-law tail an improved description of the p-p data is achieved, and the ``third component'' described in~\cite{ppprd} is eliminated. It is clear from Fig.~\ref{pipt} that below $p_t = 6$ GeV/c ($y_t = 4.5$) the true power-law trend is not easily detected.

The dotted curves intersecting the data show the two-component reference for five values of $\nu$ corresponding to the data centralities. The dashed curves bounding the dotted curves represent limiting cases $\nu = 1$, 6 for N-N and $b = 0$ Au-Au collisions respectively. The power-law trend $p_t^{-7.5}$ is illustrated. Suppression of the more-central spectra at large $p_t$ relative to the reference is qualitatively apparent. Otherwise, the detailed structure of individual spectra is better studied with more differential formats.

 \begin{figure}[h]
   \includegraphics[width=3.3in,height=3.3in]{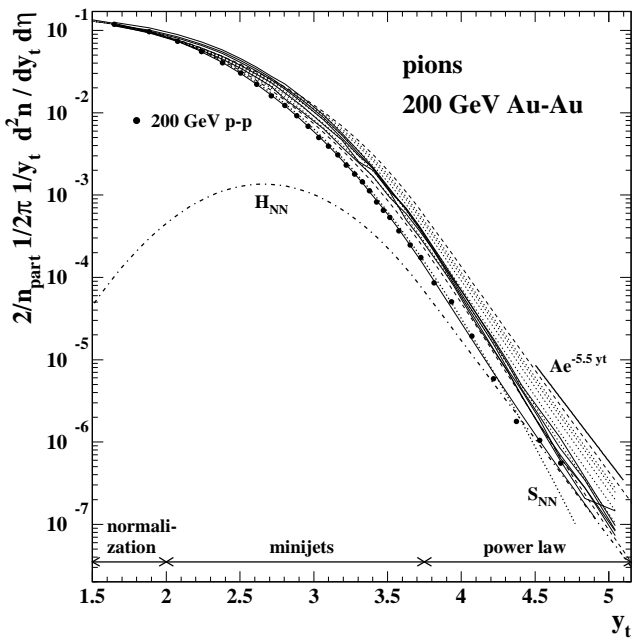}
\caption{\label{piyt}  Pion $y_t$ spectra for five Au-Au centralities (solid curves). The features are the same as for Fig.~\ref{pipt}. The power-law trend $e^{-5.5\, y_t}$ is indicated at the right side of the plot. 
 } 
 \end{figure}

In Fig.~\ref{piyt} the same spectra are plotted on $y_t$. The spectra are nearly flat at small $y_t$, aiding extrapolation and normalization. Hard-component reference $H_{NN}$ (dash-dot curve) is a Gaussian with exponential tail; the smooth transition to the power-law trend is apparent. The point of transition to the power law (in this case at $y_t = 3.75$) depends on the centroid and width of the Gaussian and the power-law exponent. The plot on $y_t$ better displays the main part of the fragment distribution centered near $y_t =$ 2.66 ($p_t \sim$ 1 GeV/c).

 \begin{figure}[h]
   \includegraphics[width=1.65in,height=1.65in]{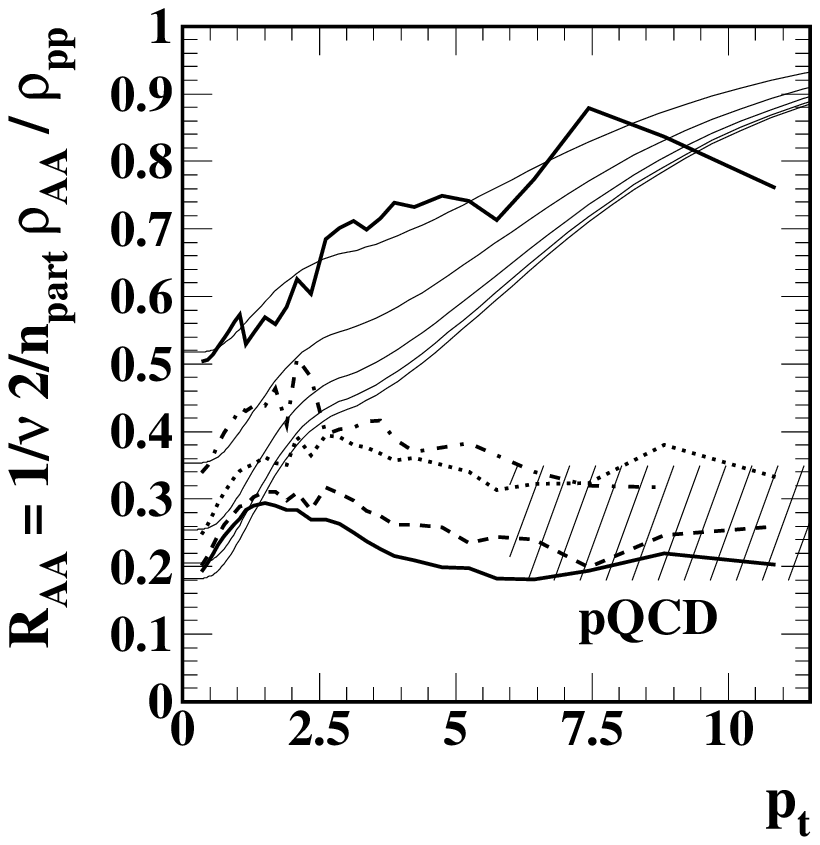}
 \includegraphics[width=1.65in,height=1.65in]{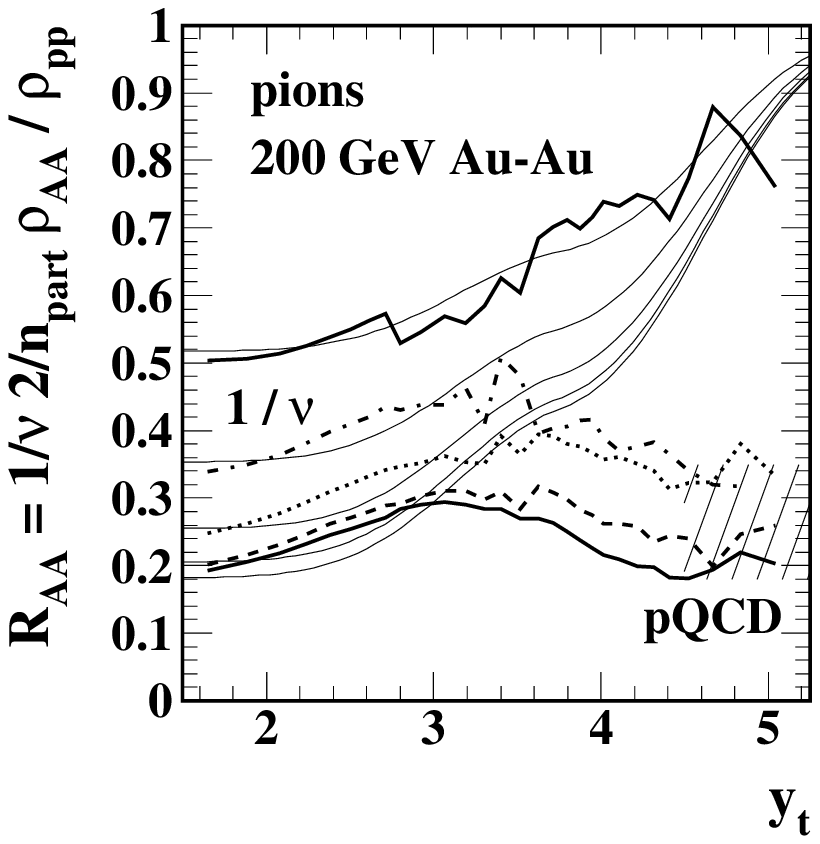}
 \caption{\label{raapions}
Left panel:  Conventional nuclear modification factor $R_{AA}$ for pions and five Au-Au centralities  (thicker curves with changing line styles). The thin reference curves are obtained from the two-component model.
Right panel: The same curves transformed to $y_t$. The $R_{AA}$ limit for small $y_t$ is $1/\nu$, with $\nu$ the mean participant path length~\cite{centmeth}.
 }
 \end{figure}
 
In Fig.~\ref{raapions} $R_{AA}$ from data (differing line styles) is plotted on $p_t$ (conventional format) and $y_t$ for comparison. The thin solid curves are the two-component references for five centralities. The main inference from $R_{AA}$ on $p_t$ (left panel) is ``jet quenching''---reduction of data {\em from unity} in the pQCD region $p_t > 6$ GeV/c. Adopting unity as a reference~\cite{star-raa} overestimates the magnitude of the reduction, since there is still a substantial soft-component contribution in that interval. The correct reference for each centrality is the corresponding thin model curve.

Turning to $R_{AA}$ plotted on $y_t$ (right panel), the data for 60-80\% (upper, bold solid curve) follow the reference within statistics, while more-central data fall below the reference on the right, but {\em exceed the reference on the left}, below $p_t \sim 2$ GeV/c. In a conventional description that part of the spectrum is also described as ``suppressed.'' The format in the left panel, with unity taken as a(n incorrect) reference, gives a false impression of parton energy loss and subsequent fragmentation. The $R_{AA}$ definition suppresses hard-component structure in the small-$p_t$ region.  The large-$p_t$ region and ``jet quenching'' are emphasized at the expense of possible new physics at smaller $y_t$. The entire fragment distribution should be compared to the correct two-component reference.

\subsection{Pion hard component} \label{pionhard}

\begin{figure}[h]
   \includegraphics[width=3.3in,height=3.3in]{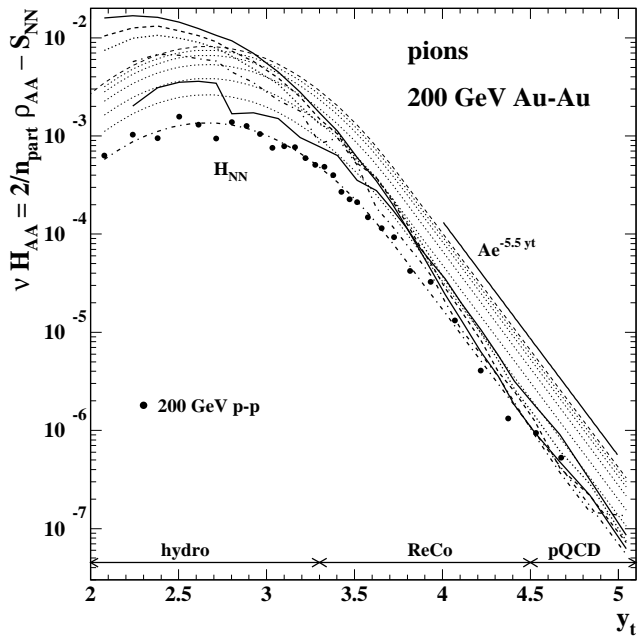}
 \caption{\label{difpi} The hard component of pion $y_t$ spectra in the form $\nu\, H_{AA}$ (thicker curves with changing line style) compared to two-component reference $\nu\, H_{NN}$ (dotted curves). The dashed reference curves are limiting cases for $\nu = 1,\,6$.
 } 
 \end{figure}

In Fig.~\ref{difpi} the $\nu \,H_{AA}$ for pions from five Au-Au centralities (bold curves of different styles) are plotted with the hard component from NSD p-p collisions~\cite{ppprd} (solid points). A common soft component $S_{NN}$ has been subtracted from the Au-Au data as in Eq.~(\ref{hard}). N-N hard-component reference $H_{NN}$ (lower, dash-dot model curve) is adopted from~\cite{ppprd} with essentially the same Gaussian parameters ($A_h = 0.333\pm0.005,\, \bar y_t = 2.66\pm0.02,\, \sigma_{y_t} = 0.45\pm0.005$). However, the hard component for this analysis also includes a power-law tail with exponential constant $n_{y_t} = 5.5\pm0.3$ (cf. App.~\ref{power}). The upper, dashed model curve is limiting case $6\,H_{NN}$ corresponding to $b = 0$ Au-Au.  The dotted reference curves between describe $\nu \, H_{NN}$ for the five data centralities, the reference system.

The agreement between $H_{NN} \sim H_{pp}$ and the solid p-p points reveals the quality of the two-component representation for elementary collisions, consistent within statistical errors. $H_{NN}$ is also statistically consistent with the 60-80\% Au-Au data for $\nu \sim 2$. In fact, the value of $n_{y_t}$ was determined by p-p and peripheral Au-Au data beyond $y_t = 4$ ($p_t = 4$ GeV/c). Incorporating the power-law tail improves the two-component description of p-p spectra. 

The remarkable new feature of this figure is the large excesses compared to the N-N reference at small $y_t$ for more central collisions and the apparent persistence of the fragment distribution there. Whereas there is a substantial reduction of fragments at large $y_t$ there is strong enhancement at small $y_t$, hinted in Fig.~\ref{raapions} (right panel) despite the $1/\nu$ suppression. That $p_t$ interval is conventionally claimed for hydro and blast-wave fitting models. However, the trend with increasing centrality is a shift to {\em smaller} $y_t$, not larger as one would expect for hydro expansion (positive $y_t$ boost). The inconsistency with hydro becomes more evident in the proton spectra.

\section{Two-component analysis -- Protons}

Identified $p_t$ spectra for protons from five centralities of 200 GeV Au-Au are shown in several plotting formats, and the hard components are isolated. The spectra were renormalized from the published data as described in App.~\ref{renorm}.  The spectra for antiprotons are similar to those for protons, but there are two significant differences, described in Sec.~\ref{protonhard}. There are major differences between pion and proton spectra, visually apparent even for $p_t$ spectra. Differences below 6 GeV/c are conventionally attributed to a combination of hydro (radial flow) and quark recombination (anomalous p/$\pi$ ratio). I examine those interpretations in the context of the two-component model in Sec.~\ref{coal}.

\subsection{Proton spectra and soft component} \label{protonspec}

\begin{figure}[h]
  \includegraphics[width=3.3in,height=3.3in]{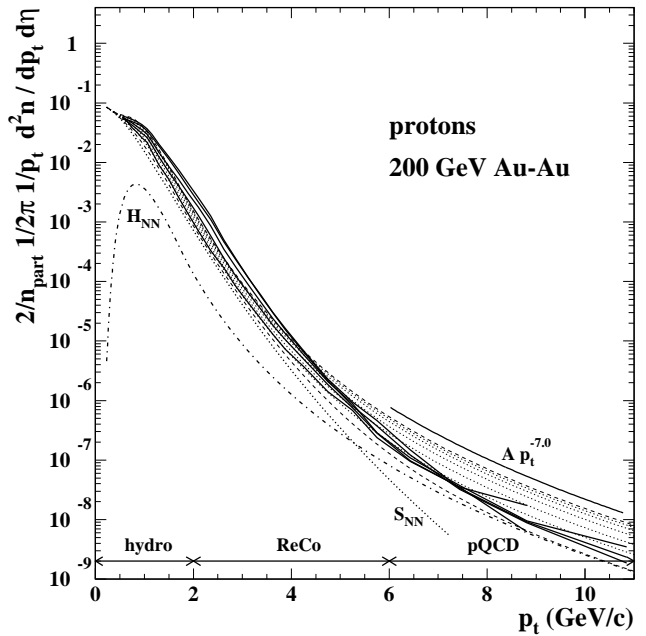}
 \caption{\label{ppt}   Proton $p_t$ spectra for five Au-Au centralities (solid curves). The general features are comparable to Fig.~\ref{pipt}. 
 } 
 \end{figure}

In Fig.~\ref{ppt}, proton $p_t$ spectra (solid curves) from~\cite{hiptspec} are plotted in the conventional format. The dotted curves are two-component reference curves for the five values of $\nu$ from data, bracketed by dashed curves for limiting cases $\nu = 1,\, 6$. Soft component $S_{NN}$ is a L\'evy distribution on proton $m_t$. The proton parameters are $A_s = 3.87\pm0.05,\, T = 0.224\pm0.005 ~\text{GeV},\, n_s = 17\pm0.2$, consistent with a previous analysis of proton spectra within $p_t \in [0.3,3]$ GeV/c~\cite{ppprd}.

As noted in Sec.~\ref{tranrap}, a substantial radial flow contribution was expected as part of the soft component, described by pion and proton Maxwell-Boltzmann distributions on $y_t$, with mass $m_0$ specific to each species but having a {\em common} positive rapidity boost $\Delta y_t$ (blast-wave model). However, the $\nu$ dependence for $y_t < 3.3$ ($p_t < 2$ GeV/c) reveals that variations with centrality are consistent with a {\em fixed} soft component nearly Maxwell-Boltzmann in form and a hard component of fixed form approximately the same as that for pions. 

 \begin{figure}[t]
  \includegraphics[width=3.3in,height=3.3in]{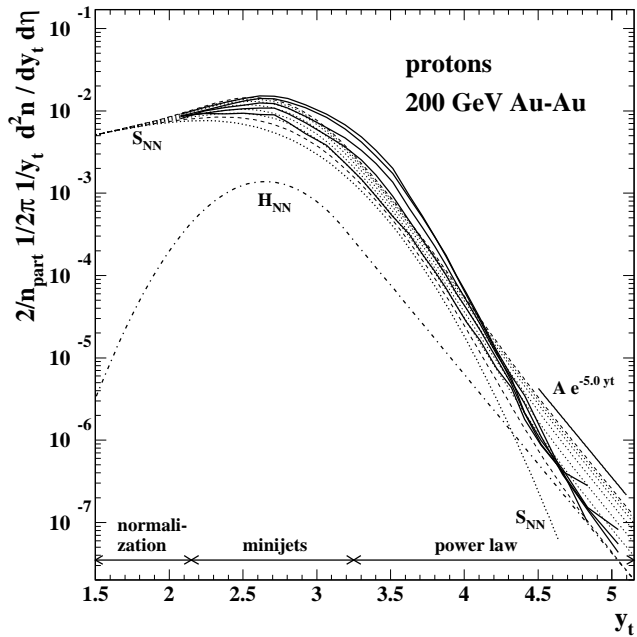}
 \caption{\label{pyt}  Proton $y_t$ spectra for five Au-Au centralities (solid curves). The general features are comparable to Fig.~\ref{piyt}. 
 } 
 \end{figure}

In Fig.~\ref{pyt} the same spectra are plotted on $y_t$ (solid curves). Again, the spectra vary slowly near $y_t = 2$, facilitating normalization and extrapolation to $p_t = 0$. The soft component $S_{NN}$ is the labeled dotted curve. The other dotted curves show the two-component reference for five values of $\nu$ matching the data, bounded by dashed curves for $\nu = 1$ and 6. At lower right a power-law trend with $n_{y_t} = 5.0$ is sketched. Dash-dot curve $H_{NN}$ is the hard-component reference.

\begin{figure}[h]
  \includegraphics[width=1.65in,height=1.65in]{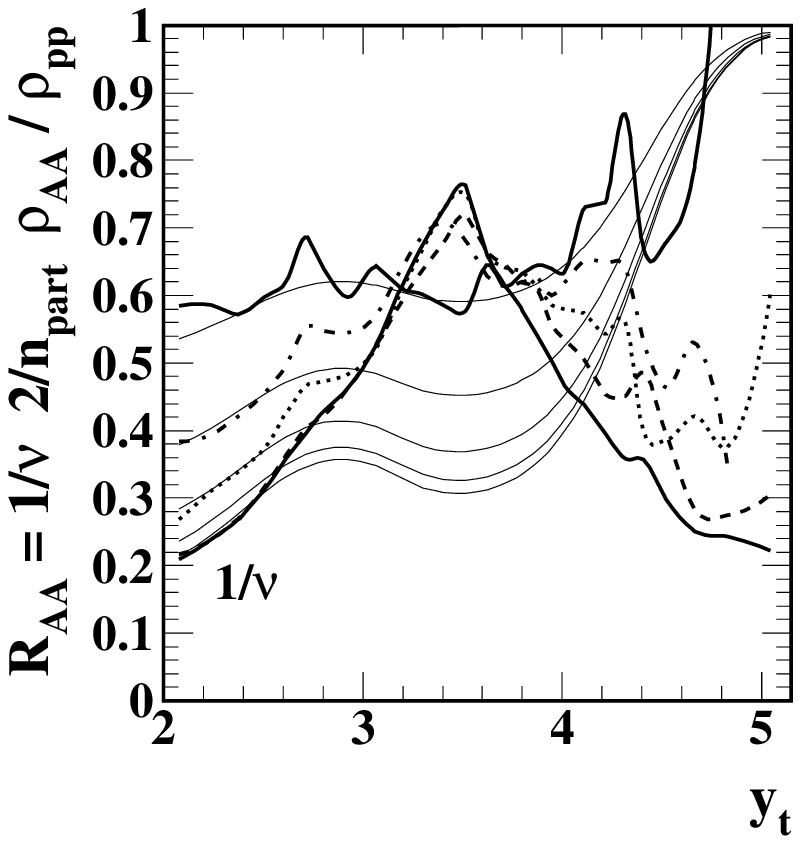}
   \includegraphics[width=1.65in,height=1.65in]{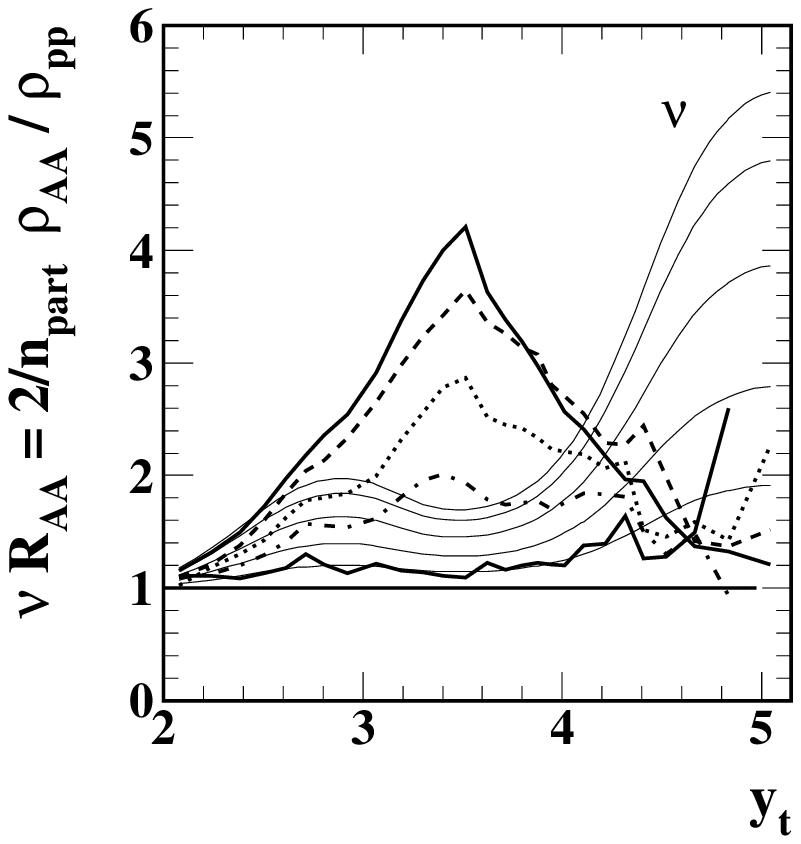}
\caption{\label{raap}
Left panel:  Conventional nuclear modification factor $R_{AA}$ for protons and five Au-Au centralities on $y_t$  (thicker curves with changing line styles). The thin reference curves are obtained from the two-component model. The $R_{AA}$ limit for small $y_t$ is $1/\nu$.
Right panel:   Alternative nuclear modification factor $\nu\, R_{AA}$ compared to thin two-component reference curves. The limit for large $y_t$ is $\nu$. The most prominent feature is a large {\em excess} near $y_t = 3.5$ ($p_t \sim 2.5$ GeV/c). 
 } 
 \end{figure}

In Fig.~\ref{raap} (left panel) conventional ratio measure $R_{AA}$ for protons is plotted on $y_t$. The apparent large suppression near $y_t = 5$ is similar to that for pions. The data in the region at intermediate $p_t \sim 2.5$ GeV/c ($y_t \sim 3.5$) are said to be ``less suppressed'' than those at larger $p_t$, indicating that the reference value is assumed to be 1 everywhere.  The thin model curves are two-component references for the five centralities, and the actual reference value for central collisions at $y_t \sim 3.5$ is $R_{AA} \sim 0.3$. Note the agreement within statistics between the 60-80\% data ($\nu \sim 2$, top-most solid data curve) and the corresponding two-component reference (top-most thin reference curve), similar to Fig.~\ref{raapions}, confirming the soft and hard reference parameters.

In Fig.~\ref{raap} (right panel) $\nu \, R_{AA}$ goes asymptotically to $\nu$ at large $y_t$. It is even more apparent in that plot that the peak near $y_t = 3.5$ is a large {\em excess} relative to the correct reference, not a suppression. A possible origin of the proton peak, the source of the so-called ``p/$\pi$ puzzle,'' is discussed in Sec.~\ref{ptopi}. Below $y_t = 2.66$ the data agree well with the two-component reference, suggesting that parton energy loss may be {\em significantly reduced for smaller parton energies}.

\subsection{Proton hard component} \label{protonhard}

\begin{figure}[h]
 \includegraphics[width=3.3in,height=3.3in]{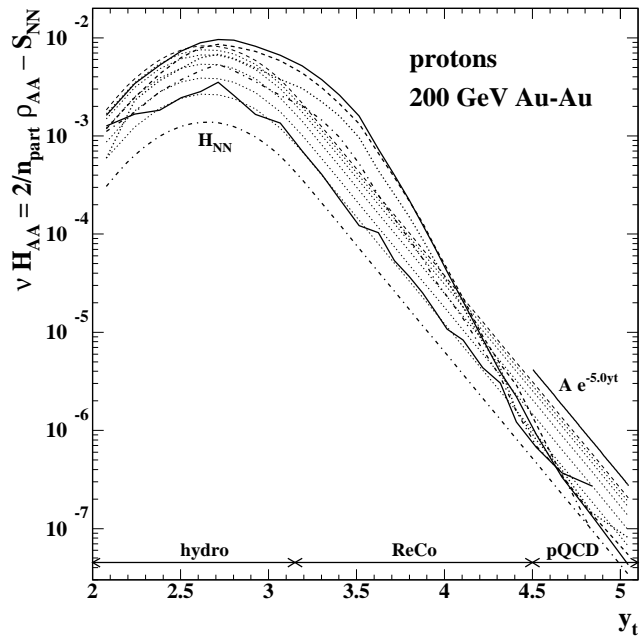}
\caption{\label{difp}  The hard component of proton $y_t$ spectra in the form $\nu\, H_{AA}$ (thicker curves with changing line style) compared to two-component reference $\nu \, H_{NN}$ (dotted curves). The general features are comparable to Fig.~\ref{difpi}. 
 } 
 \end{figure}

In Fig.~\ref{difp} I show $\nu\, H_{AA}$ for protons in five Au-Au centralities (bold curves with different line styles), and the hard-component reference $H_{NN}$ (dash-dot curve) inferred from analysis of the Au-Au spectra. There was no {\em a priori} model for the proton hard component. It was determined iteratively for this analysis just as for the p-p analysis. Spectrum variation with $\nu$ was extrapolated to $\nu = 0$ to obtain the soft-component L\'evy parameters. The $\nu$ variation of the remainder suggested the presence of a hard model function similar to that for pions and hadrons.

The proton hard-component Gaussian model is defined by $A_h = 0.475\pm0.02,\, \bar y_t = 2.66\pm0.05$ and $\sigma_{y_t} = 0.33\pm0.05$, the width substantially smaller than the pion value 0.45. The power-law index $n_{y_t} = 5.0\pm0.3$ appears to be significantly smaller (two sigma) for protons than for pions (5.5). Because of the reduced Gaussian width the proton power-law trend starts earlier, at $y_t = 3.25$. The hard-component reference with QCD power law thus plays an essential role from below 1 up to 12 GeV/c, a much broader extent than the usual pQCD description applied above 6 GeV/c. 

The dotted curves represent the two-component reference for five values of $\nu$. It is notable that the maximum values of the $H_{NN}$ for protons and pions near $y_t = 2.66$ are nearly equal. Because the proton FD is narrower the proton fragment density becomes smaller at larger $y_t$. The nearly equivalent shapes of the FDs may reflect fragmentation to different hadron species from similar underlying parton spectra, terminating just above 1 GeV/c.

Spectra for antiprotons differ from those for protons by a change in the exponent $n_{y_t}$ from 5.0 for protons to 5.5 for antiprotons (the same as for pions) and a multiplicative factor 0.8 relative to proton spectra (the factor consistent with~\cite{star2}). 

Au-Au proton data for all centralities are consistent with the two-component model below $y_t \sim 3$ ($p_t \sim 1.5$ GeV/c), providing even stronger evidence (than pions) that radial flow does not play a substantial role in spectra. The major new feature is the large excess at $y_t \sim 3.5$ for more-central collisions, which is discussed in Sec.~\ref{ptopi}.

\section{Hard-component Ratios} \label{ratios}

Hard-component references and data are compared directly in the form of ratio $r_{AA} \equiv H_{AA} / H_{NN}$ which generalizes $R_{AA}$ as a measure of parton energy loss. Whereas $R_{AA}$ is called the ``nuclear modification factor,'' $r_{AA}$ can be called simply the hard-component ratio. $R_{AA}$ suppression at large $p_t$ is ambiguous because either an attenuation factor $A< 1$ or a negative rapidity shift $\Delta y_t$ or some combination within the ``pQCD'' $y_t$ interval could result in $R_{AA} \ll 1$ for $p_t > 6$ GeV/c. In this and the next section I show that the ambiguity can be reduced by examining the structure of $r_{AA}$ over the entire $p_t$ or $y_t$ acceptance.

\subsection{Pion ratios}

\begin{figure}[h]
 \includegraphics[width=3.3in,height=3.3in]{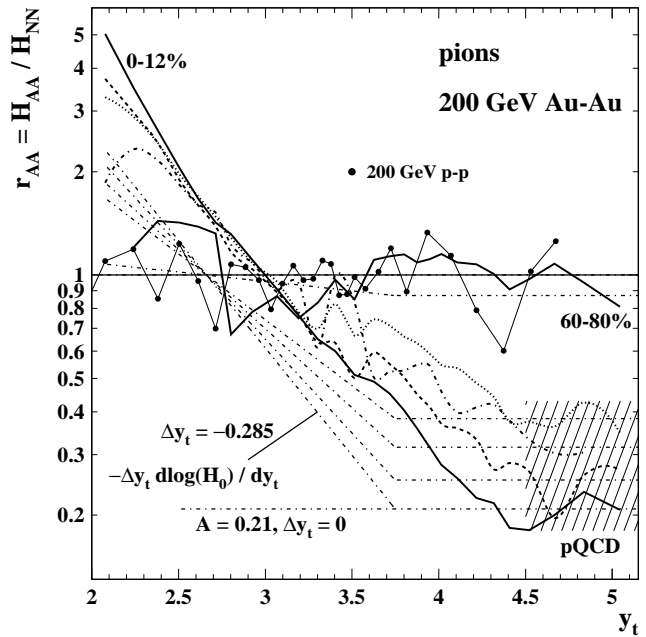}
 \caption{\label{elosspi} Hard-component ratios for pions and five Au-Au centralities (thicker curves with changing line styles) relative to the N-N hard-component reference. The connected dots are data from NSD p-p collisions. The dash-dot lines represent a simple parton energy-loss scenario (cf. Sec.~\ref{screen}). 
 } 
 \end{figure}

Fig.~\ref{elosspi} shows $H_{AA} / H_{NN}$ {\em vs} $y_t$, with $H_{AA}$ for pions from five Au-Au centrality classes (curves of several line types) and hadrons from NSD p-p collisions (solid dots). $H_{NN}$ is the model function defined in Sec.~\ref{pionhard}. The log-ratio plotting format is compatible with Eq.~(\ref{lograt}). The line $r_{AA} = 1$ describes the two-component reference for all centralities (all dotted $\nu\, H_{NN}$ curves in previous plots). Deviations from 1 represent all residuals from the two-component reference. p-p data are consistent with the model for all $n_{ch}$. The structure in this figure is then unique to heavy ion collisions. The figure can be compared with $R_{AA}$ trends in Fig.~\ref{raapions} (right panel).

Ratio data for peripheral collisions (60-80\%) are also consistent with the two-component reference. For the other four centrality classes the trend above $y_t \sim 4.5$ is consistent with $R_{AA}$ measurements in that interval (cf. Fig.~\ref{raapions}). However, below that point the data rise monotonically through 1 at $y_t \sim 3.1$ and continue to rise for smaller $y_t$, even below $y_t = 2$ ($p_t = 0.5$ GeV/c). The centrality trend near $y_t \sim 2$ is closely (anti)correlated with the trend near $y_t \sim 5$ (cf. Fig.~\ref{elosstheor} -- left panel), strongly suggesting that the two widely-separated $p_t$ regions are physically connected by the parton energy-loss and fragmentation process. More detailed centrality information for peripheral collisions is desirable to fill in the obvious gap (cf. Sec.~\ref{eloss-theor}).

That the data for peripheral collisions are consistent with p-p data and unity confirms that references $S_{NN}$ and $H_{NN}$ for Au-Au collisions are properly defined. $S_{pp}$ was defined for p-p collisions as the limiting per-particle spectrum for $n_{ch} \rightarrow 0$~\cite{ppprd}. By analogy, $S_{NN}$ should be consistent with the limiting Au-Au per-participant spectrum for $\nu \rightarrow 0$.

If a hydro contribution were present, we should expect a deviation from the two-component model at the left side of the plot, {\em shifting to the right} with increasing centrality ($\nu$). The structure we observe appears to be part of the parton energy loss process. A real hydro contribution should have a counterpart in the proton data, which we do not observe.

\subsection{Proton ratios}

\begin{figure}[t]
  \includegraphics[width=3.3in,height=3.3in]{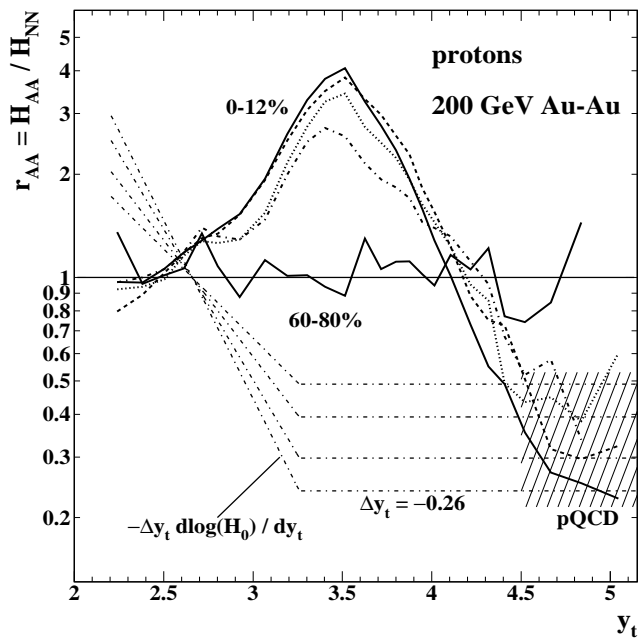}
\caption{\label{elossp}  Hard-component ratios for protons and five Au-Au centralities (thicker curves with changing line styles) relative to the N-N hard-component reference. The features are comparable to Fig.~\ref{elosspi}. 
 } 
 \end{figure}

Fig.~\ref{elossp} shows $H_{AA} / H_{NN}$ for proton data, with $H_{NN}$ defined in Sec.~\ref{protonhard}. The main differences from the pion data are the larger shift in the crossover point at unity relative to the model of Eq.~(\ref{elossmodel}) and the large excess at $y_t \sim 3.5$. This figure can be compared with $\nu \, R_{AA}$ trends in Fig.~\ref{raap} (right panel). The excess in the proton hard component {\em appears} anomalous, but may be simply explained in terms of parton energy loss.

The slopes of the proton data curves and corresponding model curves (dash-dot lines) match reasonably well as they pass through unity, although there is a horizontal shift of the crossing point (1.6 for protons, compared to 0.4 for pions). The larger proton slope at unity (compared to pions) is expected in the hypothesis of Eq.~(\ref{elossmodel}) because the width of the proton FD (and presumably the underlying proton FF for smaller parton energies) is $\sim 1.5\times$ smaller, and the $r_{AA}$ slope should therefore be $\sim 2$ times larger. The slope difference between pions and protons is thus {\em quantitatively consistent} with  the FD width difference.

The monotonic increase with decreasing $y_t$ begins near $y_t \sim 4.5$ for both pions and protons. The width difference between pion and proton FDs leads to {\em quantitatively} different evolution at smaller $y_t$ (different slopes) but may refect the same energy-loss process for both species. The return of the proton ratio to unity at $y_t \sim 2.5$ could have a counterpart in the pion spectrum at smaller $y_t$, outside the $p_t$ acceptance. The return to unity below $y_t = 2.5$, compared to the energy-loss model (sloped dash-dot lines), suggests that lower-energy partons lose relatively less (or no) energy compared to higher-energy partons. That possibility is discussed further in the next section. There is no apparent hydro structure in the proton data.

\section{Parton energy loss and color screening} \label{screen}

$R_{AA}$ data are modeled theoretically above $p_t = 6$ GeV/c ($y_t = 4.5$) where, for central Au-Au collisions at 200 GeV, the constant value is near 0.2 for pions and protons. Within the restricted $p_t$ interval it is not clear whether that value derives from a reduction of the fragment yield, a leftward shift (negative boost) of the fragment distribution, a combination or some other process. Given the ambiguity I begin with a simple model of parton energy loss as a combination of limiting cases and consider possible deviations from that reference. 

\subsection{Parton energy-loss reference} \label{frac-eloss}

A simple energy-loss model for changes in the hard component with A-A centrality is given by Eq.~(\ref{elossmodel}),
with attenuation ($A < 1$) and  left shift (boost $\Delta y_t < 0$) on $y_t$. 
Approximating the logarithm of the hard-component ratio in Eq.~(\ref{elossmodel2}) with a Taylor expansion gives
\bea \label{lograt}
\log\{H_{AA} /  H_{NN}\} \hspace{-.04in} &\simeq& \hspace{-.04in} \log(A) \hspace{-.02in} -  \hspace{-.02in} \Delta y_t\, d\log(H_{NN}) / dy_t,
\eea
with $- d\log(H_{NN}) / dy_t = (y_t - \bar y_t)/\sigma^2_{y_t}$ or $n_{y_t}$ (whichever is smaller) as obtained from the Gaussian-plus-tail model of FD $H_{NN}$ in App.~\ref{power}. Examples for $A = 1$ and $\Delta y_t$ values corresponding to the five data centralities are plotted as the dash-dot line combinations in Figs.~\ref{elosspi} and \ref{elossp}. The sloped lines pass through unity at the Gaussian centroid $y_t = 2.66$. The slopes are inversely proportional to the FD width squared. For the most central collisions in Fig.~\ref{elosspi} alternative limiting case $A \sim 0.21 $ with $\Delta y_t = 0$ is indicated by the labeled extension and seems not to describe the data. The negative-boost energy-loss model ($A = 1,\, \Delta y_t < 0$) indicates the form of the ratio resulting from shifting  the {\em entire} FD down on rapidity with no other changes. The negative-boost model does not describe the data quantitatively, but its form provides a useful reference.


The relative $p_t$ reduction corresponding to boost $\Delta y_t < 0$ is given by
\bea
p'_t \approx m_0 \, \exp\{y_t + \Delta y_t\}/2  &\approx&  e^{\Delta y_t} \, p_t \\ \nonumber
{\Delta p_t}/{p_t} &\approx& \Delta y_t.
\eea
From  Figs.~\ref{elosspi} and~\ref{elossp} $\Delta y_t \sim -0.25$ for pion and proton spectra above $p_t = 6$ GeV/c in central Au-Au collisions, implying that about 25\% of the parton (and fragment) momentum is ``lost'' in central Au-Au collisions {\em for energetic partons}. It is not clear from the data whether parton energy is actually lost to a separate ``medium'' or instead is {\em rearranged} within the fragment distribution. The energy loss for lower-energy partons may be much less (cf. Sec.~\ref{screen-eloss}). The seemingly large fragment attenuation at larger $p_t$ is apparently the consequence of a large power-law exponent and a small rapidity shift.

\subsection{$r_{AA}$ modeling and centrality evolution}

\begin{figure}[h]
  \includegraphics[width=3.3in,height=1.65in]{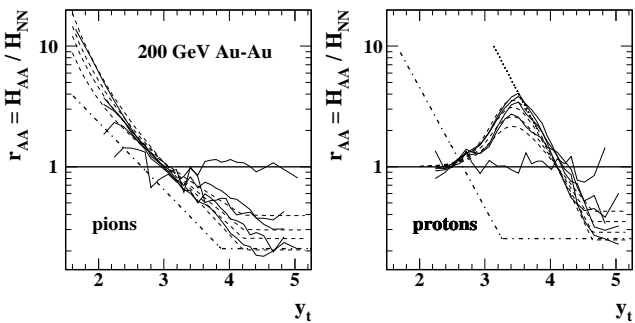}
\caption{\label{elossmod}  
Left panel: Hard-component ratios $r_{AA}$ for pions from Au-Au collisions for five centralities (solid curves). The dash-dot lines represent a rapidity shift $\Delta y_t$ (negative boost or red shift) of the entire N-N hard-component reference $H_{NN}$ for 0-12\% central collisions. The dashed curves are $r_{AA}$ model functions defined in the text.
Right panel: Hard-component ratio $r_{AA}$ for protons from Au-Au collisions (solid curve). The features are similar to the left panel. 
 } 
 \end{figure}

In Fig.~\ref{elossmod} hard-component ratios $r_{AA}$ are plotted {\em vs} $y_t$ for pions (left panel) and protons (right panel) from Au-Au collisions in five centrality classes. The dash-dot lines represent the FD negative-boost energy-loss scenario---uniform shift of the entire hard-component FD to smaller rapidity---for central collisions. Despite the horizontal offset between data and model curves the similarities suggest that parton energy loss is dominated by a negative boost of {\em some part} of the underlying parton distribution. The offsets suggest that lower-energy partons experience less energy loss. 

The dashed curves in Fig.~\ref{elossmod} which closely model the $r_{AA}$ data are defined by
\bea \label{raapis}
\log[r_{AA}(y_t;\nu)]   &=&   - \Delta y_t(\nu)\, d\log(H'_{NN}) / dy_t+ \\ \nonumber
&& 23\exp\{-1.9\,y_t\}  
\eea
for pions and
\bea \label{raaprotons}
\log[r_{AA}(y_t;\nu)]  &=&   -\Delta y_t(\nu)\, d\log(H'_{NN}) / dy_t \times \\ \nonumber
&& \hspace{-.39in} \{\tanh([y_t -3.24]/0.37) + 1\}/2  
\eea
for protons. $\Delta y_t(\nu)$ is the negative boost for path-length (centrality) $\nu$, and $H'_{NN}$ is $H_{NN}$ with centroid shifted from 2.66 to 3.1 for pions and 4.2 for protons. The $r_{AA}$ models assume for simplicity that all centrality dependence is contained in multiplicative factor $\Delta y_t$, the values shown as points in Fig.~\ref{elosstheor} (left panel). The small difference between pion and proton $r_{AA}$ amplitudes at $y_t \sim 5$ may be due to the different exponents $n_{y_t} = 5.5,\, 5.0$ for pions and protons respectively.

The {\em ad hoc} $\tanh$ factor in Eq.~(\ref{raaprotons}) is well-determined by the proton data. However, the additional term in Eq.~(\ref{raapis}) is uncertain below $y_t \sim 2$ where significant additional fragment yield appears. That uncertainty has implications for pion integrals and $\langle p_t \rangle$ estimates.

\subsection{Comparison with energy-loss theory} \label{eloss-theor}

Parton energy loss in RHIC collisions is of major theoretical interest~\cite{elosstheor1,elosstheor2,vitev}. I select one example~\cite{vitev} to provide a comparison with this analysis.

 \begin{figure}[h]
  \includegraphics[width=1.65in,height=1.65in]{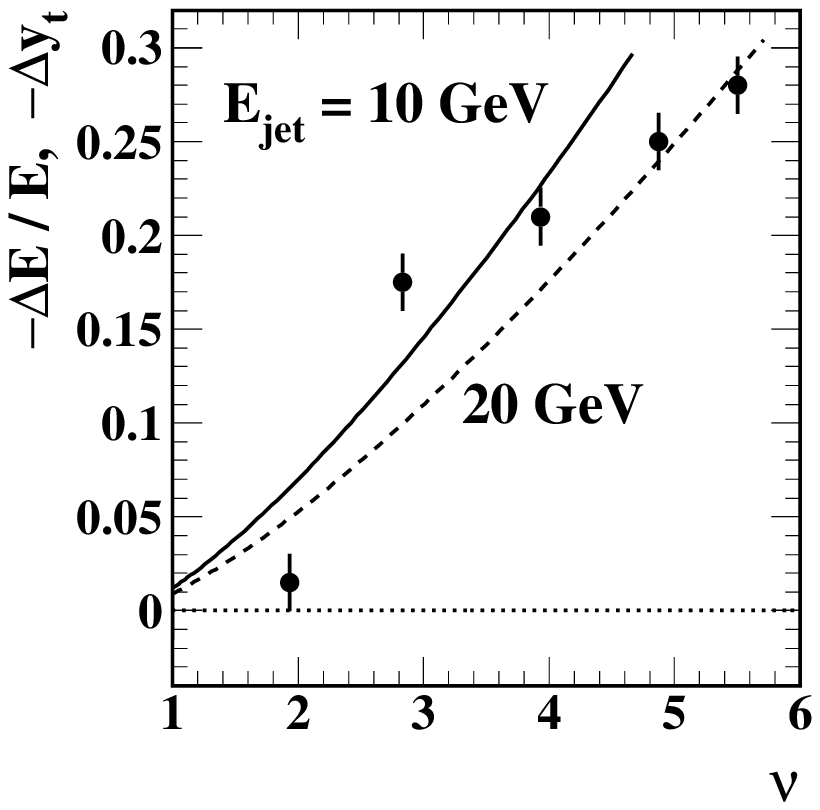}
  \includegraphics[width=1.65in,height=1.61in]{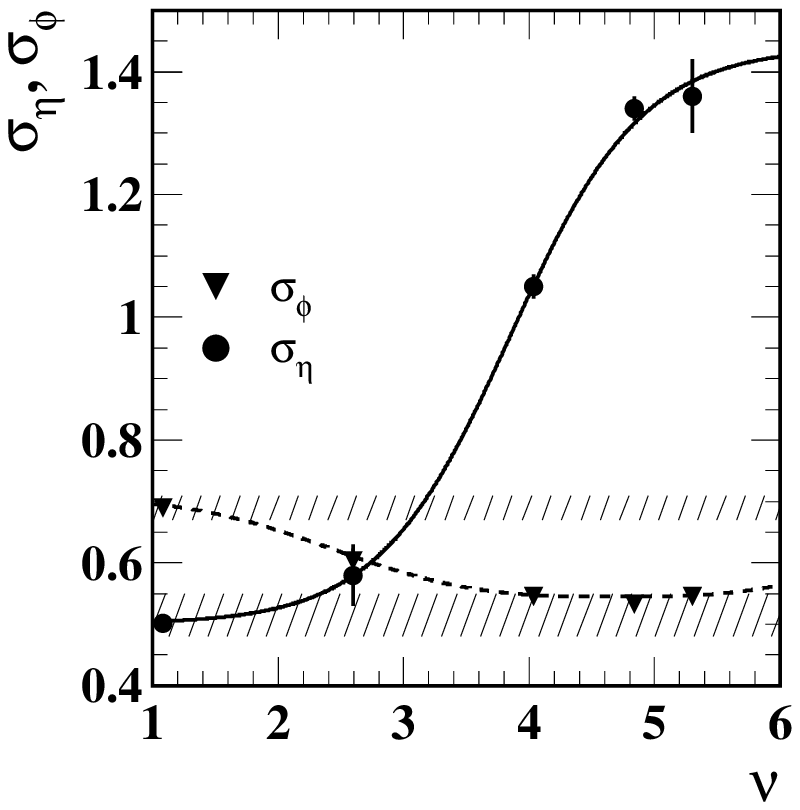}
\caption{\label{elosstheor}  
Left panel: Negative boosts $\Delta y_t$ for $y_t$ spectra from five centralities of Au-Au collisions at 200 GeV (points) and theoretical predictions of relative energy loss $\Delta E / E$ for two parton energies (curves).
Right panel: Same-side minijet peak widths from four centralities of Au-Au collisions at 130 GeV compared to values for p-p collisions ($\nu = 1$, hatched regions).  
 } 
 \end{figure}

In Fig.~\ref{elosstheor} (left panel) I show values of $\Delta y_t$ obtained from this analysis {\em vs} participant path length $\nu$. As noted in Sec.~\ref{frac-eloss}, $\Delta y_t \sim \Delta p_t / p_t \sim \Delta E / E$ (signed numbers) for parton transverse momentum and energy. I also plot predicted relative energy loss $\Delta E / E$ for 10 GeV (solid curve) and 20 GeV (dashed curve) partons {\em vs} centrality. 

Two issues arise from this comparison. First, the data trend appears to exhibit a step-wise transition near $\nu = 2.5$, already apparent in Figs.~\ref{elosspi} and \ref{elossp}. The data above the transition vary approximately as $-\Delta y_t \sim 0.05\,(\nu + 0.5)$. The theoretical centrality dependence is $\propto n_{part}^{2/3}$~\cite{vitev}. Because $\nu \sim (n_{part}/2)^{1/3}$ the prediction is therefore $\Delta E /E \propto \nu^2$, roughly gluon density ($\sim \nu$) $\times$ outgoing parton pathlength ($\sim \nu$). The predicted magnitudes are close to the data, but there is no discontinuity in the theory. A step structure is also seen in minijet correlation peak systematics. Fig.~\ref{elosstheor} (right panel) shows same-side minijet peak widths from 130 GeV Au-Au collisions ($p_t \in [0.15,2]$ GeV/c), in which the $\eta$ width increases dramatically from p-p values (hatched regions) above $\nu = 3$~\cite{axialci}. 

Second, the energy dependence of relative energy loss is described as monotonically decreasing above $E_{jet} \sim 10$ GeV (Fig.~2 of~\cite{vitev}). That trend would be indicated by return of $r_{AA}$ to 1 with {\em increasing} $y_t$, which is not observed within the data $p_t$ acceptance. At smaller $y_t$ the trend of $r_{AA}$ for both pions and protons is a return to (and through) unity, which could be interpreted as a decrease in relative parton energy loss, an increase in fragment number due to energy loss or both. The proton data especially seem to indicate that low-energy partons lose little or no energy (cf. Sec.~\ref{screen-eloss}).

For reference, at the transition point $\nu \sim 2.5$ (average number of N-N collisions per participant pair) there are on average about 19 participant nucleon pairs and 48 binary N-N collisions (given $\sigma_{NN} = 42$ mb), compared to 191 and 1136 respectively for $b = 0$ Au-Au. The fractional centrality is 58\% (100\% = N-N) and $b = 11$ fm. 

\subsection{Color screening and charge screening} \label{screen-eloss}

The structure of Figs.~\ref{elosspi} and~\ref{elossp} compared to the reference suggest that partons with smaller energies experience less fractional energy loss compared to larger energies. For instance, the hard component of proton spectra near $p_t \sim 1$ GeV/c is consistent with the p-p hard component, {\em even for central Au-Au collisions}, whereas the FD negative boost above 6 GeV/c is consistent with a 25\% parton energy loss. That interpretation, if correct, suggests the presence of {\em color screening} for partons---partons become increasingly ``white'' at smaller energy scales. 

Electric charge screening is observed in the energy loss of charged particles, especially heavy ions passing through crystalline media in certain orientations (channeling). The mean ion charge depends on the ion speed in such a way that straggling is reduced: more energetic ions carry larger exposed charges and lose more energy per unit path length, and conversely. Charge attachment from the medium to slower ions is more probable, leading to projectile charge screening. The consequent reduction of energy straggling is described as ``bunching''~\cite{bunching}. A similar screening phenomenon in RHIC collisions could lead to the observed enhancement of minijet structure relative to the N-N reference.

Energy-scale-dependent color screening is consistent with $e^+$-$e^-$ jet multiplicity trends. The distinction between gluon and quark jet multiplicities, nominally determined by Casimir color factors $\text{C}_\text{F}$ and $\text{C}_\text{A}$, vanishes below $Q = 10$ GeV (5 GeV partons)~\cite{lepmini}. Even at $\sqrt{s} \sim 90$ GeV the gluon-to-quark jet multiplicity ratio $n_g / n_q \sim 1.4$ is significantly below $\text{C}_\text{A}/\text{C}_\text{F} = 2.25$~\cite{abreu}. That value (QCD) and 1 (non-QCD) are used to estimate parton energy loss in~\cite{elosstheor2}, where the effect on $R_{AA}$ of differing color charge is predicted to {\em increase} with decreasing fragment momentum.

\section{The proton-to-pion ratio} \label{ptopi}

\begin{figure}[h]
  \includegraphics[width=3.3in,height=1.65in]{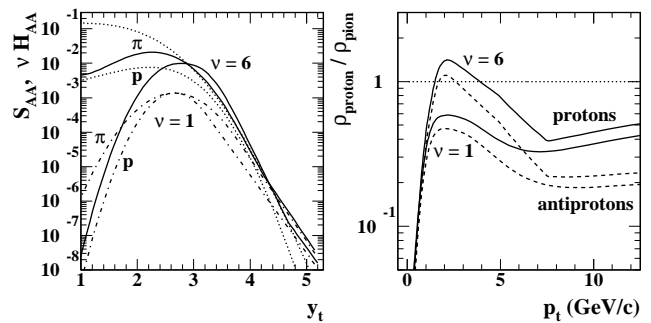}
\caption{\label{ppirat}  
Left panel: Model functions inferred from this analysis. Dotted curves are soft components for protons and pions. Dash-dot curves are hard-component N-N ($\nu = 1$) references. The solid curves are hard components for $\nu = 6$ generated from the hard-component N-N references plus parton energy-loss parameterizations $r_{AA}(y_t;\nu)$.
Right panel: Proton-to-pion two-component full-spectrum ratios on $p_t$. The plot can be compared to Fig.~4 of~\cite{hiptspec}.
} 
 \end{figure}

We can use the two-component spectrum model to study the proton-to-pion ratio, which has received considerable theoretical attention~\cite{hwarec,friesrec,grecrec}. Fig.~\ref{ppirat} (left panel) summarizes the model functions for pions and protons obtained from this analysis. The dotted curves are the fixed soft components $S_{NN}$. The dash-dot curves are the hard components $H_{NN}$ for $\nu = 1$ (N-N collisions), and the solid curves are $H_{AA}$ for $\nu \sim 6$  ($b = 0$ Au-Au).

Using the expressions for $r_{AA}(y_t;\nu)$ obtained in the previous section densities $2/n_{part}\, \rho_{AAh} = \{ S_{NN} + \nu\, r_{AA}(y_t;\nu) H_{NN}\}$ are defined for the two hadron species. Spectrum ratios $\rho_\text{AAproton} / \rho_\text{AApion}$ are plotted in the right panel, labeled ``protons'' for $\nu = 1,\, 6$ (solid curves). The ratios for antiprotons (dashed curves) are obtained (according to observed data properties) by changing $n_{y_t}$ from 5.0 for protons to 5.5 (the pion value) and multiplying the resulting proton density by 0.8 to obtain the antiproton density. Those results can be compared with Fig.~4 of~\cite{hiptspec}. Correspondence with the measured ratios is very good. The model ratios from this analysis are slightly larger due to the spectrum renormalization. The proton-to-pion puzzle is thus apparently transformed to details of parton energy loss and modified fragmentation.

The full-spectrum ratios share the property of $R_{AA}$ that they mix soft and hard spectrum components, suppressing details at smaller $p_t$. The change of the spectrum ratio with centrality is actually modest. There is at the peak (2 GeV/c) only a factor $2\times$ increase for central Au-Au collisions relative to N-N collisions. In the left panel the hard-component ratio for N-N collisions (dash-dot curves) is 1 already at $y_t = 2.66$ ($p_t \sim 1$ GeV/c), descends to 0.5 near 4 GeV/c and then {\em rises through unity again} due to the apparent difference in the underlying parton spectra (that for protons being harder). For central Au-Au collisions we observe an excess of protons above the FD mode and an {\em even larger excess} of pions below the mode. The differential hard components provide a much more detailed story than full-spectrum ratios.

Comparing to ratios of fragmentation {\em functions} (FFs) from $e^+$-$e^-$ collisions introduces further confusion because the spectrum hard components are fragment {\em distributions} (FDs)---integrals of FFs folded with parton spectra. Even given that distinction, theory fragmentation functions such as the KKP parameterization typically describe only the 10\% most energetic fragments in an FF and seriously diverge from fragmentation data below that point~\cite{lepmini}, where A-A spectrum issues are most complex. In general, fully-differential formats such as Figs.~\ref{elosspi} and \ref{elossp} provide a clearer comparison of hadron species.

\section{Coalescence and Recombination} \label{coal}

Ratios $r_{AA}$ are an improvement on $R_{AA}$ because they isolate hard components from soft, but absolute magnitudes are lost. Additional insight is gained by studying transport of absolute quantities, particularly transverse mass, to test measure conservation during fragmentation or coalescence (competing models of hadron production). 

The two-component reference is an initial-state model: some hadrons are produced by immediate fragmentation of participant nucleons, some from large-angle parton scattering out of those nucleons and subsequent fragmentation. Coalescence models describe hadron production by combinations of pairs and triples of partons both within and between the two {\em parton} spectrum components. Additional production of particles, momenta or quantum numbers relative to the reference could indicate extraordinary transport from longitudinal to transverse phase space.

\subsection{Hard-component Differences}

\begin{figure}[h]
  \includegraphics[width=3.3in,height=3.3in]{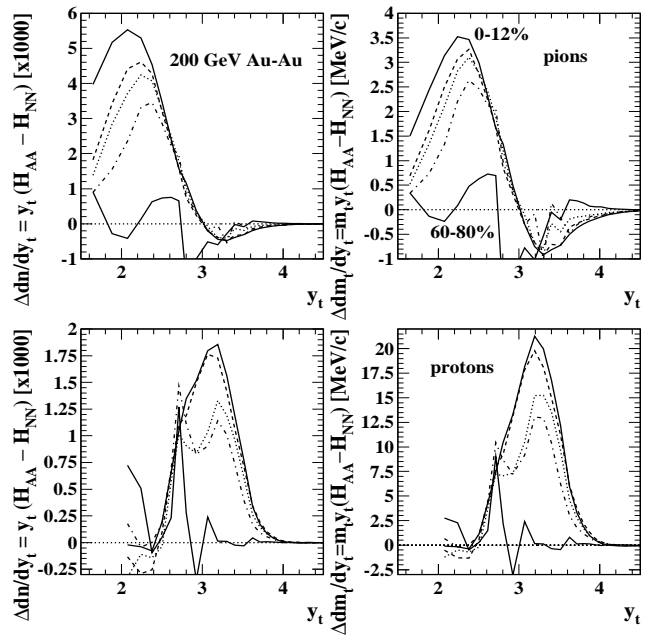}
\caption{\label{harddif}  
Left panels: Number transport of pions (upper panel) and protons (lower panel) for five Au-Au centralities relative to hard-component reference $H_{NN}$.
Right panels: $m_t$ transport of pions (upper panel) and protons (lower panel) for five Au-Au centralities relative to hard-component reference $H_{NN}$.
 } 
 \end{figure}

In Fig.~\ref{harddif} particle number transport (left panels) and $m_t$ transport (right panels) relative to two-component references are plotted. 
There are two significant deviations from the reference: 1) particle (and $m_t$) suppression at larger $p_t$ and 2) particle (and $m_t$) excess at smaller $p_t$. Because suppression at larger $p_t$ involves a small number of particles it is not visually apparent for protons and is a subtle feature for pions in this format. The conventional explanation for 1) is parton energy loss; that for 2) is coalescence/recombination. This analysis suggests that given their closely correlated centrality dependence the two features may have a common origin.

\subsection{Coalescence and Recombination}

Recombination was invoked for RHIC data in response to anomalous values of the proton-to-pion ratio at intermediate $p_t \in [2,6]$ GeV/c for central Au-Au collisions~\cite{hwarec,grecrec,friesrec}. The theoretical approach accounts for {\em reduced suppression} of protons in that interval, compared to the region at larger $p_t$ described by pQCD, by parton coalescence or recombination.

The recombination mechanism assumes the presence of two partonic components, a thermalized, flowing partonic medium (QGP) with approximately exponential parton spectrum dominating at smaller $p_t$, and a spectrum of hard-scattered partons described by a fragmentation function (e.g., KKP) or a power-law function dominating at larger $p_t$. The magnitudes of the two parton spectrum components cross over within the intermediate $p_t$ interval. A coupling of the two components is proposed through parton coalescence to form hadrons, resulting in additional hadron yield in the intermediate $p_t$ region compared to spectra from elementary collisions. Differences in spectrum details lead to the p/$\pi$ ratio anomaly.

\subsection{Assumed spectrum shapes}

Given the present heavy-ion and previous p-p two-component analysis there are problems with the spectrum components invoked in the coalescence model. Assuming that the soft component in elementary collisions (partons and hadrons) is already thermal (including radial flow) is unjustified. The hard component measured in N-N collisions, a fragment distribution built on a true power-law function (cf. App.~\ref{power}), is different from the fragmentation function (e.g., KKP) or conventional power-law function (modified L\'evy distribution) assumed in the theoretical models. The differences at larger $p_t$ are very significant. Results based on assumed spectrum features may be qualitatively different from data.

The p-p two-component spectrum decomposition~\cite{ppprd} revealed two independent {\em non-thermal} components resulting from fragmentation in two orthogonal directions. The notion that the soft component is the result of thermalization should be contrasted with the possibility that the transverse soft component in N-N collisions is hadron fragments from the longitudinal ``hard'' component---partons from the participant nucleon PDF. At what point on Au-Au collision centrality does that situation change?

A critical issue for the ReCo model is the locations of the spectrum crossovers which determine where the enhancement by coalescence should appear. In~\cite{hwarec,grecrec,friesrec} the pion crossover is in 2-4 GeV/c and the proton crossover is in 4-6 GeV/c. The result for the p/$\pi$ ratio is a peak within 2-4 GeV/c as shown in Fig.~\ref{ppirat} (right panel). 
However, in this analysis we find that contrary to the coalescence picture the ratio trend results from a combination of different hard-component widths plus the proton excess centered at 2 GeV/c (not 4-6 GeV/c). The general shape of the ratio distribution is already established for N-N collisions where thermalization and QGP are not at issue, and the ratio increase with centrality near 2 GeV/c is due entirely to the localized proton excess, apparently related to parton energy loss.

\subsection{Two-particle correlations}

Two-particle correlations present additional challenges for recombination. Even for the soft component in peripheral Au-Au (N-N) collisions we observe strong angular correlations suggesting a one-to-two (parton-to-hadrons) process leading to strongly charge- and momentum-correlated hadron pairs. We also observe strongly-correlated hadrons in the hard component down to 0.35 GeV/c which exhibit all the characteristics of jet correlations arising from fragmentation and which join smoothly onto the systematics of trigger-particle jet correlations as the selected hadron $p_t$ is increased.

Thermal partons are assumed to be uncorrelated in the recombination model~\cite{hwarec}, in which case it is not clear how to obtain the hadron correlations observed in the soft component of N-N collisions. For the hard component, hadrons are modeled as SS or SSS combinations of shower partons. Back-to-back hadron pairs in the away-side jet would require a correlated combination of four, five or six shower partons from two jets to generate the hadron correlations, requiring a degree of coordination among partons not explained in current coalescence models.

\subsection{Summary}

To be competitive, recombination/coalescence models must describe all aspects of the differential hadron spectra, including both hard and soft spectrum components, not just integrals or selected ratios. Whereas there is an unusually large baryon/meson ratio in one $p_t$ interval, there is an unusually large meson/baryon ratio in another (not predicted by present ReCo models). Both must be described by theory. Also, such models should transition smoothly from {\em in vacuo} N-N collisions (where there is no ``thermalization'') to central Au-Au collisions (where there may or may not be any ``thermalization'').

The picture that emerges from this analysis is one in which parton fragmentation remains the dominant hadronization mechanism in all $p_t$ intervals, but interactions among partons become important for larger densities (centralities). The hard components are strongly modified, but still maintain their integrity as parton fragment distributions. The differences between pion and proton FDs are substantial, but may simply be due to quantitative differences in the underlying FFs. Recombination and coalescence models should be tested against that detailed experimental context.

\section{Integral spectrum measures} \label{partprod}

The two-component model with $r_{AA}(y_t;\nu)$ from this analysis describes spectrum data to their statistical limits. The model functions can be integrated to determine the centrality systematics of individual components and total spectra and compared to minimum-bias multiplicity distributions.

\subsection{Spectrum models}

The two-component models of pion and proton spectra from this study (3D densities) are summarized by
\bea  \label{summ}
\frac{2}{n_{part}} \rho_\pi \hspace{-.04in} &=& \hspace{-.04in} \frac{0.85\, \rho_0}{1.012} \{ S_{0\pi}  \hspace{-.02in}+ \hspace{-.02in} 0.012\, \nu\, r_{AA\pi}(y_t;\nu)\,  H_{0\pi} \} \\ \nonumber
\frac{2}{n_{part}} \rho_p \hspace{-.04in} &=& \hspace{-.04in} \frac{0.062\, \rho_0}{1.118} \{ S_{0p} \hspace{-.02in} + \hspace{-.02in} 0.118\, \nu\, r_{AAp}(y_t;\nu)\,  H_{0p} \},
\eea
with $\rho_0 = 2.5 / 2 \pi$. The unit-integral model functions and hard-component ratios $r_{AA}$ are defined above. The $r_{AA}$ represent all deviations from the two-component reference. If we set $r_{AA} = 1$ and integrate Eqs.~(\ref{summ}) on $y_t$ the two-component constant for pions is $x = 0.012$, whereas the K-N description requires $x_{KN} = 0.1$ for the hadron yield at 200 GeV~\cite{nardi}. The apparent contradiction must be contained in the $r_{AA}$, which is confirmed here quantitatively.

\subsection{Spectrum integrals}

\begin{figure}[h]
  \includegraphics[width=3.3in,height=1.65in]{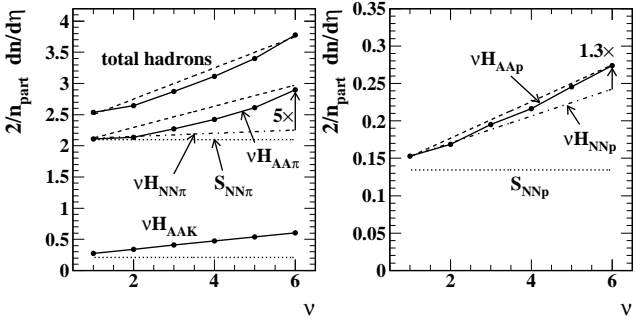}
 \caption{\label{norm1}
Left panel: Spectrum integrals for pion soft component (upper dotted line) and hard component $\int dy_t\,y_t\, H_{AA\pi}$ which sum to the middle solid curve. The dash-dot line extrapolates the pion hard-component reference from N-N collisions. The total and soft-component kaon contributions (lower solid curve and dotted line) are rough estimates. The upper solid curve is the total yield $\pi + K + p$.
Right panel:  Similar curves for integrated proton spectra. The dash-dot line extrapolates the proton hard-component reference from N-N collisions. The differences between solid and dash-dot curves for pions and protons result from parton energy-loss factors $r_{AA}\neq 1$. 
 } 
 \end{figure}

In Fig.~\ref{norm1} (left panel) I show the integral of the charged pion (both signs) spectrum [first line of Eq.~(\ref{summ})] evaluated at six values of $\nu$ (middle solid curve and points). The symbols refer to the integrated densities. The dotted line is the integral of pion soft component $S_{NN\pi}$. The solid curve corresponds to the $r_{AA}$ from data, including parton energy loss and resulting fragmentation changes. The dash-dot line $0.85\,\rho_{0} [1 + 0.012(\nu - 1)]$ is the integral if $r_{AA} = 1$ in the spectrum model, describing the extrapolation from N-N collisions. 

Parton energy loss is thus responsible for a factor 5$\times$ increase in pion fragments for central Au-Au collisions relative to the N-N extrapolation on $\nu$. As noted previously, there is a substantial uncertainty in the pion $r_{AA}$ below $y_t \sim 2$ which produces a comparable systematic uncertainty in the hard-component integral for more central collisions. The relative uncertainty in the pion hard component may be as large as 50\%.  The corresponding relative error in the pion solid curve in Fig.~\ref{norm1} is 20\%.

In Fig.~\ref{norm1} (right panel) I show equivalent results for the proton spectrum (particle plus antiparticle) model. The proton hard component in N-N collisions is a much larger fraction of the total proton yield (11\% compared to 1.2\% for pions). The increase of fragment yield from parton energy loss is only 30\% for protons (from dash-dot to solid curve) compared to factor 5$\times$ for pions.

The lowest curves in the left panel provide a rough estimate of the kaon yield and centrality dependence. Adding kaons to the pion and proton results gives the top solid curve in the left panel, which is in fair agreement with the K-N two-component expression for the per-participant density of unidentified hadrons at 200 GeV, $2.5\,[1 + 0.1\,(\nu - 1)]$ (top-most dashed line). We can now understand the large hard-component excess in Au-Au collisions over the N-N extrapolation. The main effect is the substantial increase in pion yield at small $p_t$ resulting from parton energy loss. The parton fragment yield for central Au-Au collisions is apparently 1/3 of the total multiplicity, and those fragments remain strongly correlated.

\subsection{$\langle p_t \rangle$ systematics}

We can also use the model spectra to determine $\langle p_t \rangle$, the ensemble mean $p_t$ defined by
\bea \label{mean-pt}
 \rho_{0}\, \langle p_t(\nu) \rangle &=& {\int_0^\infty dp_t\, p^2_t\, \rho(p_t;\nu) }
 \\ \nonumber
&=& \rho_{s}\,\langle p_{t} \rangle_{S_{NN}} + \nu \rho_{h}\, \langle p_{t}(\nu) \rangle_{H_{AA}},
\eea
with the densities scaled together appropritiately for each hadron species abundance. In the two-component model the mean $p_t$ also has two components, as in the second line of Eq.~(\ref{mean-pt}): a soft component independent of centrality and a hard component.

 \begin{figure}[h]
  \includegraphics[width=3.3in,height=1.65in]{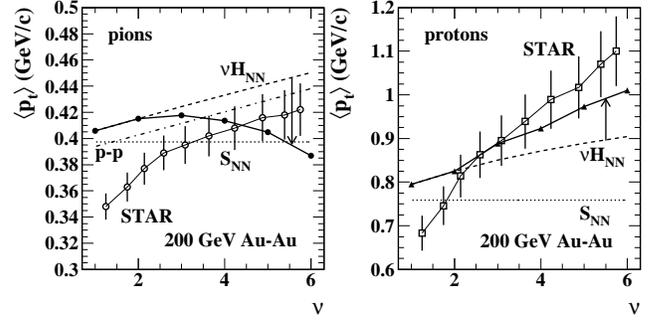}
 \caption{\label{meanpt}
Left panel: Pion $\langle p_t \rangle$ centrality trends from two-component model functions. The dotted line is the fixed soft component. The dashed line adds the extrapolated N-N reference $\nu H_{NN}$. The solid curve with points is the complete model with $r_{AA}(y_t,\nu)$. The open symbols are measured means from extrapolated $p_t$ spectra~\cite{star}. The dash-dot line is the two-component trend from p-p analysis assuming $n_{ch} / \bar n_{ch} \leftrightarrow \nu$~\cite{ppprd}.
Right panel: Same as the left panel, but for proton spectra.
 } 
 \end{figure}

Fig.~\ref{meanpt} shows $\langle p_t \rangle$ values obtained with the model functions from this analysis. The dotted lines indicate the constant soft component $\langle p_t \rangle_{S_{NN}}$. The dashed lines give the two-component trend expected for no parton energy loss:  $r_{AA} \rightarrow 1$ and $\nu H_{AA} \rightarrow  \nu H_{NN}$ with constant $\langle p_t\rangle_{H_{NN}}$. The solid curves and points are from the full data model with energy loss and evolving $\langle p_{t}(\nu) \rangle_{H_{AA}}$. $\rho_s$ and $\rho_h$ are given in Eq.~(\ref{summ}). The soft- and hard-component reference means are $\langle p_t\rangle_{S_{NN}} = 0.40\pm 0.02,\, 0.76\pm0.03$ GeV/c and $\langle p_t\rangle_{H_{NN}} = 1.19\pm 0.01,\, 1.12\pm 0.01$ GeV/c, for pions and protons respectively in each case.

For pions (left panel) the spectrum mean (solid curve) can drop below the dashed line (arrow) because the additional pion yield for central collisions appears to the left of the $H_{NN}$ mode, as shown in Fig.~\ref{difpi}. However, the solid curve should be considered a lower limit (cf. Sec.~\ref{unchard} for further discussion). For protons (right panel) the solid curve rises above the dashed line (arrow) because the additional proton yield relative to the reference hard component appears to the right of the $H_{NN}$ mode in Fig.~\ref{difp}. 

The dash-dot line (left panel) summarizes the p-p multiplicity dependence of $\langle p_t \rangle$ from~\cite{ppprd} assuming $n_{ch} / \bar n_{ch} \leftrightarrow \nu$. The results are consistent within the 0.02 GeV/c systematic uncertainty in extrapolating the spectrum soft component to zero common to all $n_{ch}$ or $\nu$~\cite{ppprd}.
Also shown are mean values extracted from STAR spectra (open symbols) extrapolated from more-limited $p_t$ intervals: [0.2,0.7] GeV/c for negative pions and [0.5,1.05] GeV/c for antiprotons~\cite{star}. 

The apparent mass dependence of $\langle p_t \rangle$ and increase with A-A centrality is commonly interpreted to result from radial flow in heavy ion collisions, described for instance by a blast-wave model. Flow velocities are then inferred from spectrum fits~\cite{star}. From the present analysis the major contribution to the mass dependence is the soft component which does not change with A-A centrality. Most of the mass dependence is already present in N-N collisions.  Support for an inferred ``radial-flow'' velocity $\langle \beta \rangle$ is derived from the nearly linear increase of minijet yields, peaked near $p_t =$ 1 GeV/c, with $\nu \sim (n_{part}/2)^{1/3}$. That dependence leads to the characteristic $(dn/d\eta)^{1/3}$ dependence of some inferred $\langle \beta \rangle$ centrality trends.

\section{Model Uncertainties}

This spectrum decomposition is based on three assumptions. 1) Deviations from participant scaling are minimal at small $p_t$ or $y_t$ (the hard component is small there). That region can be used to normalize measured spectra. 2) The first-order deviation from participant scaling and soft component $S_0(y_t)$ is the reference hard component $\propto \nu\, H_{0}(y_t)$. Thus, $S_0(y_t)$ and $H_0(y_t)$ model functions are determined by variations with $\nu$ in the limit $\nu \rightarrow 0$. 3) Any nonlinearity at larger $\nu$ is represented by factor $r_{AA}(y_t;\nu)$ in the hard component. Those assumptions are expressed by Eq.~(\ref{aa2comp}), and the limit process is defined in principle by
\bea \label{2complimit}
\rho_h H_0(y_t) &\equiv& \stackrel{\nu \rightarrow 0}{\lim}  \{2/n_{part}\, d\rho_{AA}(y_t;\nu)/d\nu\}  \\ \nonumber
\rho_s S_0(y_t) &\equiv& \stackrel{\nu \rightarrow 0}{\lim}  \{2/n_{part}\, \rho_{AA}(y_t;\nu) - \nu \rho_h \, H_0(y_t)\}.
\eea
In practice, the relation between spectra for two peripheral centralities and/or p-p collisions is used for the initial determination of $S_0$ and $H_0$, and the two-component parameters are refined once the entire system with differential plots is established. For each hadron species in this analysis the model parameters are the five spectrum normalizations, two shape parameters for each of $S_0$ and $H_0$ plus exponential (power-law) parameter $n_{y_t}$ and the hard/soft relative abundance $ \bar x = \rho_h / \rho_s$. 

\subsection{Spectrum normalization}

Complete {\em a priori} ignorance of measured spectrum normalization was assumed, based on conclusions in App.~\ref{renorm}. Initial guesses were based on the K-N two-component description of particle production measured by minimum-bias multiplicity distributions. The 200 GeV p-p spectrum is normalized to 2D density $\rho_0 = 2.5/2\pi$. Au-Au $p_t$ spectra are normalized to $\rho_0$ for $\nu = 1$, and the relative normalizations for other $\nu$ are initially determined by the K-N trend. Relative normalizations are refined after the two-component analysis is established. Consequences of normalization uncertainties are negligible, as argued below.

While assumption 1) (the low $p_t/y_t$ region is slow-varying with multiplicity or centrality) is true for hadrons in p-p and protons in Au-Au collisions, it is not true for pions in Au-Au collisions (because of parton energy loss). However, the pion soft component falls rapidly while the hard component rises rapidly in the region near $y_t = 2$. A 1\% change in spectrum normalization produces large changes in hard components below $y_t \sim 2.5$ ($H_{NN}$ is about 1\% of $S_{NN}$ at $y_t \sim 2$, cf. Fig.~\ref{difpi}).  Requiring that the inferred hard component vary slowly with $y_t$ in that region reduces the relative normalization uncertainty below 2\% for $y_t > 2.5$.

The form of the proton spectrum on $y_t$ is such that normalization uncertainty plays a negligible role in model uncertainties. Below $y_t \sim 3$ the spectra exhibit near-ideal two-component behavior. The soft component is nearly flat where the hard component has significant magnitude. Applying the K-N normalization to the proton spectra according to App.~\ref{renorm} already results in convergence to a single $S_{NN}$ for small $y_t$. Asymptotic approach of spectra to the soft component as $y_t \rightarrow 1.5$ is clear in Fig.~\ref{pyt}. 

\subsection{Soft component}

In general, the procedure of Eq.~(\ref{2complimit}) is applied to define $S_{NN}(y_t) = \rho_s\, S_0(y_t)$ and $H_{NN}(y_t) = \rho_h\, H_0(y_t)$. For instance, spectra for $\nu = 1,\,2$ would form ``simultaneous equations'' for the two functions. Functions $S_{pph}$ and $H_{pph}$ were obtained previously for hadrons in p-p collisions where we observe ideal two-component behavior~\cite{ppprd}. Thus, for pions $S_{NN\pi}$ was initially assigned the parameters $(A_s,\,T,\,n_s)$ = (20.3,\,0.1445 GeV,\,12.8) from $S_{pph}$. The result was a small but significant disagreement between inferred hard component $H_{NN\pi}$ and $H_{pph}$. By reducing $n_s$ to 12.0 consistency was achieved. The required change in $n_s$ was attributed to the species mix of the hadron spectrum from p-p collisions and lack of a power-law tail in the hard-component model. The soft parameter uncertainties are then similar to those in~\cite{ppprd}. 

No initial values were assumed for proton $S_{NNp}$ or $H_{NNp}$; model parameters were determined solely by the iterative process defined by Eq.~(\ref{2complimit}).  The final soft parameters from this analysis for $p_t \in [0.5,12]$ GeV/c are  $(A_s,\,T,\,n_s)$ = (5.25,\,0.223 GeV,\,17), consistent with the proton soft component inferred from spectrum data for $p_t \in [0.3,3]$ GeV/c~\cite{star2} (cf. Sec.~IX of~\cite{ppprd}). 

\subsection{Hard component} \label{unchard}

The hard spectrum components and their ratios to reference distributions are the main result of this analysis. Sources of relative error for the hard components are 1) normalization error relative to fixed soft reference $S_{NN}$, common to all centralities (substantial for pions, mainly below $y_t = 2.5$), 2) errors in relative normalizations of different centralities (negligible in all cases), 3) errors in $S_{NN}$ shape parameters, mainly $n_s$ (changes the $H_{AA}$ shape above the mode).

Referring to Fig.~\ref{difpi}, reduction of the overall pion normalization by 1\% produces a 50\% reduction in the $\nu \sim 2$ (peripheral) centrality at $y_t \sim 2.25$. The change for $\nu \sim 5.5$ (central) is negligible. Normalization issues are most critical for peripheral centralities where the hard component is small. Associated changes at $y_t \sim 3$ are negligible.

Changes in $n_s = 12$ by $\pm 0.5$ can be compensated at larger $y_t$ by adjusting the normalization of the soft component by factor $1\mp 0.05$. However, the structure at smaller $y_t$ is then strongly distorted. In particular, the agreement between $\nu \sim 2$ data and the reference (unity) in Fig.~\ref{elosspi} is spoiled at small $y_t$. 

Given an optimized soft component, increase of the $H_{NN}$ centroid by 0.05 causes a 20\% decrease of the $\nu \sim 2$ data above the centroid and 20\% increase below the centroid in Fig.~\ref{elosspi}. A decrease in the width $0.445 \rightarrow 0.435$ causes an 20\% increase of $\nu \sim 2$ data above $y_t = 3$.

Systematic variations for protons are similar. The  60-80\% (most peripheral) data provide a strong constraint on the fixed model parameters. Evolution with centrality in Figs.~\ref{elosspi} and \ref{elossp} is then a characteristic of the data, not of the fixed model. The system of curves in those figures can be shifted or distorted {\em collectively} to some degree by adjusting the soft model parameters, but not their relative spacings.

The $\langle p_t \rangle$ determination for pions is quite uncertain for more central collisions because the shape of $r_{AA}$ is not defined by data below $y_t \sim 2$ or $p_t \sim 0.5$ GeV/c, and the fragment yield in that region increases by a large factor ($\sim 5\times$) with centrality. The effect of the additional fragment yield on the spectrum mean depends on its location relative to 0.4 GeV/c (the soft-component mean) which is not determined by data. Thus, in Fig.~\ref{meanpt} (left panel) the solid curve plus points should be interpreted as a lower-limit estimate, and the dashed line an upper-limit estimate. In contrast, the proton $\langle p_t \rangle$ (right panel) relative to the soft mean is well-defined for all centralities.
 
\subsection{Further comments on uncertainties}

It can be argued that the two-component K-N normalization trends were imposed on the analysis at the beginning, so the results in Fig.~\ref{norm1} are circular. However, as discussed above the normalization constraints from the data themselves actually determine the final normalizations. That the K-N trends are approximately in agreement is then a consistency check, an indication of small systematic error. The more important lesson of Fig.~\ref{norm1} is the demonstration of soft- and hard-component contributions to the spectrum integral, their variation with centrality, and the dramatic role of parton energy loss in particle (especially pion) production in central Au-Au collisions. 

Figs.~\ref{elosspi} and \ref{elossp} show the residual differences between data and reference which contain information about parton energy loss. The 20\%  systematic uncertainty estimates (upper limits) in the absolute values of the data ratios should be compared with excursions on $y_t$ in those figures from 1/5 to greater than 10, a 50-fold variation. The relative structure in those figures is therefore very significant compared to the model uncertainties. If the two-component model is viewed as a ``fit'' to the data by the reference then Figs.~\ref{elosspi} and \ref{elossp} show all the fit residuals and reveal details of parton energy loss for all $p_t/y_t$ in the acceptance.

\section{Discussion}

Conventional interpretation of $p_t$ spectra from central heavy ion collisions is predicated on the formation of a thermalized and flowing bulk medium which strongly dissipates energetic partons. The present study, a differential analysis assuming only linear superposition of two functional forms, reveals a different picture with surprising simplicity. The hard component, all of which may result from parton fragmentation, extends well below 1 GeV/c and dominates the spectrum centrality evolution. 

\subsection{Two-component analysis}

The supporting context for two-component analysis of Au-Au  $p_t$ spectra is provided by soft and hard components in elementary collisions: $e^+$-$e^-$ fragmentation functions described over the entire fragment distribution~\cite{lepmini}, p-p two-component analysis~\cite{ppprd} and extensive correlation analysis revealing minijet structure~\cite{axialci,mtxmt,ptscale,edep}. The question for heavy ion collisions is how the two components evolve with A-A centrality, and what that evolution tells us about QCD. The present analysis reveals new information on parton stopping and fragmentation and provides critical tests of current theoretical models.

\subsection{The soft component}

The fixed form of soft component $S_0$ is formally defined in this and the p-p two-component analysis~\cite{ppprd} as the limiting spectrum shape for $\nu \rightarrow 0$ or $x \rightarrow 0$ respectively. As such, there is no physical model invoked. Particle production arises from an unspecified soft process (longitudinal fragmentation of individual participant or ``wounded'' nucleons). By hypothesis, the soft component does not change form with A-A centrality. Does some of the structure in Figs.~\ref{elosspi} and~\ref{elossp} actually belong to a changing soft component? The structure at large $y_t$ is consistent with pQCD expectations, and the structure at smaller $y_t$ is strongly correlated with it for pions and protons. The results of data analysis appear consistent with two-component assumptions concerning $S_0$.

\subsection{The hard component}

Hard-component model $H_{NN}(y_t)$ has a simple fixed form consistent with pQCD for larger $y_t$, with recent results from p-p spectra for smaller $y_t$~\cite{ppprd}, and with $e^+$-$e^-$ fragmentation functions~\cite{lepmini}. It appears to be a minimum-bias fragment distribution, the folding of an unbiased parton spectrum with conditional fragmentation functions, extending down to $0.3$ GeV/c.

The Gaussian parameters of the Au-Au pion hard component are little changed from those inferred from p-p hadron spectra. There was no {\em a priori} information about the proton hard component. The form was inferred in this analysis from the centrality ($\nu$) dependence of Au-Au spectra following the iterative method applied to $n_{ch}$ dependence of p-p spectra in~\cite{ppprd}. The $H_{NN}$ distributions are similar: Gaussians on $y_t$ with exponential tails, having the same centroids but different widths and QCD exponents. The proton width is about 2/3 the pion width. 

\subsection{Parton energy loss and color screening}

Measurements of parton energy loss inferred from nuclear modification of parton fragmentation in the form $R_{AA}$ are expected to reveal properties of the QCD medium produced in RHIC heavy ion collisions. Hard-component ratio $r_{AA}$ measured over the entire $p_t$ or $y_t$ acceptance provides much more information, given the limited range of validity of $R_{AA}$. In contrast to a picture of ``jet quenching'' as absorption of partons in an opaque medium, the present analysis suggests that no partons are ``lost'' in A-A collisions. Their manifestation (in spectrum structure and correlations) is simply redistributed within the fragment momentum distribution, and the fragment number increases.  A high-$p_t$ triggered jet yield may be reduced by a factor of five {\em within particular $p_t$ cuts}, but additional fragments emerge elsewhere, still with jet-like correlation structure~\cite{mtxmt,axialci}. 

Figs.~\ref{elosspi} and~\ref{elossp} provide the first clear indication from spectra of how parton energy is {\em transported} within the fragmentation process. Much of the difference between hard-component ratios $r_{AA}$ for pions and protons may derive from the differing widths of the elementary (N-N) conditional fragmentation functions. The common indication from both species is an apparent {decrease} of relative parton energy loss with smaller parton energy, suggesting screening of color charge. 

\subsection{The Cronin effect}

The Cronin effect is a modification of the $p_t$ spectrum in p-A collisions relative to a Glauber linear superposition of p-p collisions~\cite{cronin}. The observed effect is measured by nuclear modification factor $R_{pA}$ and consists of an excess ($R_{pA} > 1$) in the $p_t$ interval [2,\,6] GeV/c. The Cronin effect is conventionally modeled as initial-state parton multiple scattering leading to increased $k_t$. However, it was suggested recently that the Cronin effect could be a final-state effect due to recombination~\cite{hwa}.

Cronin enhancement has been contrasted with suppression of ratio $R_{AA}$ observed at RHIC for $p_t  > 6$ GeV/c and attributed to jet quenching. Enhancement or suppression is measured relative to reference value $R = 1$. What we learn from Figs.~\ref{raapions} and~\ref{raap} is that the true reference for such ratios is not unity. What has been called suppression may well be {\em enhancement relative to the correct reference}, and conversely. In Figs.~\ref{raap} and~\ref{elossp} we see a large proton enhancement for central Au-Au collisions in just the $p_t$ region ($y_t \in [3,4.5]$) where the Cronin enhancement is observed, and with a similar form. It is possible therefore that at least part of the Cronin effect is a final-state phenomenon related to parton energy loss, albeit in cold nuclear matter.

\subsection{$y_t$ spectra and minijets}

The conventional RHIC picture of thermalized Au-Au collisions at 200 GeV is contradicted by the abundance of surviving minijets (fragments from a minimum-bias parton spectrum) observed in two-particle angular and momentum correlations~\cite{axialci,ptscale,edep}. The ``minijet excess'' directly conflicts with claims of near-ideal hydrodynamics and ``perfect liquid.'' Two-particle momentum correlations on $m_t \times m_t$ ($y_t \in [1.3,3]$) apparently revealed the lower half of the pion $H_{AA}$ moving down on transverse momentum with increasing Au-Au centrality~\cite{mtxmt} (cf.~the systematics of Fig.~\ref{difpi}). Two-particle angular number correlations reveal strong ``minijet deformation'' in the same $p_t$ range~\cite{axialci}. The energy and centrality systematics of two-particle angular $p_t$ correlations reveal the source mechanism of $\langle p_t \rangle$ fluctuations and provide new details of parton-medium interactions~\cite{ptscale,edep}.

The present analysis appears to resolve the contradiction. Although large-momentum partons do suffer substantial energy loss ($\sim$ 25\%), small-momentum partons apparently do not. Single-particle $p_t$ spectra in this analysis are consistent with minijet manifestations in two-particle correlations which imply that low-energy partons are not absorbed by the medium. They are modified to some degree, but much jet structure remains at smaller $p_t$ or $y_t$. Color screening would explain the large abundance of minijet correlations observed in central Au-Au collisions.

\subsection{Further implications from this analysis}

Conventional spectrum analysis isolates different intervals of the $p_t$ spectrum. $R_{AA}$ emphasizes larger $p_t$ and visually suppresses the strong minijet contribution at smaller $p_t$, which is then reinterpreted in terms of soft physics models (e.g., radial flow).  Spectra are fitted with monolithic functions motivated by thermodynamic, statistical and hydro models whose parameters may not be meaningful but which are interpreted to reveal a structureless, thermalized, flowing bulk medium. 

In contrast, the two-component spectrum model describes several physical mechanisms consistently over the entire $p_t$ acceptance. Instead of a thermalized medium we find that 2/3 of the hadrons in central Au-Au collisions are part of a soft component apparently unchanged from N-N collisions, and 1/3 are part of a hard component which appears to be strongly-correlated parton fragments consistent with QCD expectations.  In that context the ``p/$\pi$ puzzle'' is also a manifestation of parton energy loss.

Since 1/3 of the hadrons originate effectively from rapidly-moving sources (parton  fragmentation), the significance of statistical-model spectrum measures (chemical potentials, decoupling temperatures, $\langle p_t \rangle$s) attributed to an expanding bulk medium can be strongly questioned. Upon close examination of pion and proton spectra no identifiable radial flow is apparent. Inference of an equation of state or phase transition also seems problematic. The best evidence for a QCD medium may come from high-statistics minijet correlations~\cite{axialci,mtxmt,ptscale,edep}.

 \section{Summary}

In this paper I develop two-component model functions to describe $p_t$ and $y_t$ spectra for identified pions and protons extending to 12 GeV/c. Spectra for five centralities from Au-Au collisions at 200 GeV are accurately represented to the statistical limits of the data by a hard+soft model. This analysis extends a similar study of $p_t$ spectra for unidentified hadrons from p-p collisions at 200 GeV which provided reference data for the present analysis.

The soft component has the form of a L\'evy distribution on transverse mass $m_t$ as in p-p collisions and appears to be independent of centrality. The hard-component reference is  a Gaussian on transverse rapidity $y_t$, with exponential tail corresponding to the expected QCD power-law trend $1/p_t^{n_h}$ and required to describe $p_t$ spectra beyond 6 GeV/c. 
The Gaussian width for protons is smaller than for pions by factor 2/3, but the peak modes are the same. 

The centrality dependence of the hard-component {\em reference} is $\propto \nu = 2 n_{bin} / n_{part}$, reflecting binary collision scaling. Evolution of the {\em data} hard component with centrality relative to the reference is represented by ratio $r_{AA}$ which generalizes nuclear modification factor $R_{AA}$. Whereas the latter is a ratio of total $p_t$ spectra, including  soft components, the former is a ratio of isolated hard components (data/reference). As such, $r_{AA}$ reveals {\em without distortion} the centrality evolution of the complete parton fragment distribution over all $p_t$ or $y_t$ in the acceptance---all ``parton energy loss'' in Au-Au collisions.

The evolution of $r_{AA}$ with centrality---the main result of this paper---is simply described. Parton energy loss produces a shift or negative boost $\Delta y_t$ of {\em part} of the minimum-bias fragment distribution (Gaussian plus tail). Much of the difference in $r_{AA}$ structure for protons and pions may result from the substantial width difference of the N-N hard components. $r_{AA}$ structure at smaller $y_t$ suggests that energy loss of low-energy partons is small, and color screening may be important at smaller energy scales. Two-component spectrum analysis thus provides access to new aspects of parton energy loss in A-A collisions over the complete parton spectrum  and resolves several open questions from minijet correlation studies.

This work was supported in part by the Office of Science of the U.S. DoE under grant DE-FG03-97ER41020

\begin{appendix}

\section{Hard-component model} \label{power}

$p_t$ spectra are expected to go asymptotically to power-law trend $1/p_t^{n}$ at larger $p_t$, reflecting an underlying power-law parton spectrum. In~\cite{ppprd} it was demonstrated that the hard component in p-p $p_t$ spectra is well described by a Gaussian on $y_t$ for $p_t < 6$ GeV/c. However, that description is inadequate at larger $p_t$, and a power-law trend must be restored to the model. The need is already apparent in Fig.~10 (left panel) of~\cite{ppprd} where the data for $\hat n_{ch} = 11.5$ rise above the soft L\'evy plus hard Gaussian model function for $p_t > 4$ GeV/c. In  this appendix I describe how an exponential tail can be added to the Gaussian on $y_t$ to restore the QCD power law to the hard-component model.

\subsection{Algebraic description}

The hard-component model function is defined on transverse rapidity $y_t$. If the QCD power law is $\rho(p_t)  \propto p_t^{-n_h}$ and
\bea
\rho(y_t) &=& \frac{m_t\, p_t}{y_t} \rho(p_t) \\ \nonumber
\text{then}~~~ -d\log[\rho(y_t)] / dy_t &\sim& (n_h-2) + 1/y_t
\eea
Since the region relevant to the power-law trend is $y_t \sim 5$, and systematic uncertainties in the exponent are comparable to 0.2, I define $n_{y_t} = n_h - 2$ as the relevant exponential constant on $y_t$. Since $n_h \sim 7.5$ I expect $n_{y_t} \sim 5.5$ for data.

 \begin{figure}[h]
   \includegraphics[width=3.3in,height=1.58in]{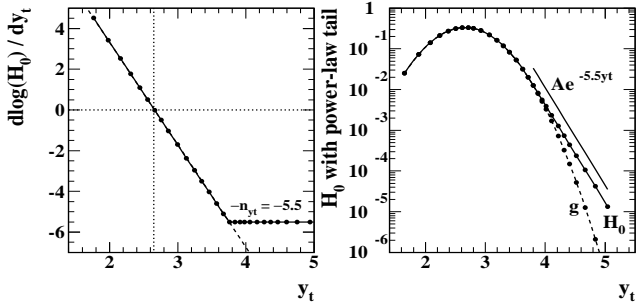}
\ \caption{\label{gpower}
Left panel: The logarithm derivative for a Gaussian (dashed line) and Gaussian plus exponential tail (solid lines).
Right panel:  Gaussian (dashed curve) and Gaussian plus exponential (power-law) tail (solid curve). The points correspond to $y_t$ values for the data used in this analysis. 
 } 
 \end{figure}

In Fig.~\ref{gpower} the algebraic strategy is illustrated. The hard-component Gaussian model from~\cite{ppprd} (dashed curve $g$ in the right panel) is 
\bea
g(y_t) = A_h\, \exp\{-[(y_t - \bar y_t)/\sigma_{y_t}]^2/2\},
\eea
with logarithm derivative $d\log(g) / dy_t = -(y_t - \bar y_t) / \sigma^2_{y_t}$ shown by the dashed line in the left panel. To add an exponential tail to the Gaussian the logarithm derivative in the left panel must be limited from below by fixed value $- n_{y_t}$ (solid line in the left panel). The running integral of the resulting function (solid curve in the right panel) is then exponentiated to obtain the desired Gaussian with exponential tail as hard-component model $H_0(y_t)$.

\subsection{Numerical algorithm}

To illustrate a discrete numerical analysis I assume $n$ arbitrarily-spaced data points at  $ y_{t,i}$ shown by the dots in Fig.~\ref{gpower} (the $y_t$ values from this analysis). Define abbreviations $lg \equiv  \log(g)$, $lH_0  \equiv \log(H_0)$, $lg' = dlg/dy_t = -(y_t - \bar y_t) / \sigma^2_{y_t}$, $lH'_{0} = d\, l H_0/dy_{t}$

Obtain $lg_i = \log[g(y_{t,i})]$ \\
(right panel, points on dashed curve)

Obtain forward differences $dx:~ dx_{i+1} = x_{i+1} - x_{i}$, with $dx_{1} = 0$, for $x = y_t$, $lg$

Obtain logarithm derivative $lg'_i = dlg_i/dy_{t,i} $ 

Obtain forward average $\bar{lg'}_{i+1} \equiv (lg'_{i+1} + lg'_i)/2$

Obtain forward average $\bar y_{t,i+1} \equiv (y_{t,i+1} + y_{t,i})/2$ 

Apply condition to produce exponential tail
\bea
 lH'_{0,i} &=&  \bar{lg'_i}~~~ \text{if} ~~~ \bar{lg'_i} > - n_y \\ \nonumber
 &= &  -n_{y_t}  ~~~ \text{if} ~~~ \bar{lg'_i} \leq - n_{y_t}
\eea
(left panel, points on solid lines)

Obtain running integral
\bea
lH_{0,j} = \sum_{i=1}^{j}  lH'_{0,i} \cdot dy_{ti} + lg_1~~~j \in [1,n]
\eea \\
(right panel, points on solid curve).
The hard-component model function with exponential tail is then $H_{0,i} =A_h\, \exp(lH_{0,i})$, with $A_h$ determined by the unit normal condition.

Extension of the hard-component model to a Gaussian with exponential tail (power law on $p_t$) results in qualitatively better descriptions of data beyond $p_t = 6$ GeV/c. Description of p-p and peripheral Au-Au collisions is very good to 12 GeV/c. The QCD power-law exponent is expected to be $n_h \sim 7.5$, or $n_{y_t} \sim 5.5$. There is a possible difference between pion and proton exponents---7.5 {\em vs} 7.0 respectively---at a two-sigma level of significance.

\section{Spectrum Normalization} \label{renorm}

\begin{figure}[h]
  \includegraphics[width=3.3in,height=1.65in]{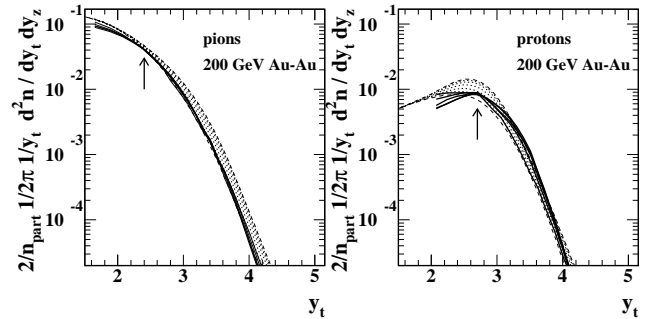}
 \caption{\label{nonorm} Published spectra (solid curves) for pions and protons for five Au-Au centralities with original normalization~\cite{hiptspec}. The arrows indicate where the spectra cross over. Dashed and dotted curves show the final two-component reference system from this analysis.
 } 
 \end{figure}

Fig.~\ref{nonorm} shows published data from~\cite{hiptspec} without renormalization (solid curves) compared to the final two-component model from this analysis (dashed and dotted curves). Published spectra were used in the form $n_{\pi^+} + n_{\pi^-}$, $n_p$ and $n_{\bar p}$. When integrated, the spectra should lead to total yields whose centrality systematics agree with minimum-bias distributions, a basic constraint on spectrum normalization. 

The curve crossings at $y_t \sim 2.5 - 3$ (arrows) suggest that there is a normalization problem. As published, the data indicate that per-participant proton and pion yields below $y_t \sim 2.5$ ($p_t \sim 1$ GeV/c) {\em decrease} with increasing Au-Au centrality. 
The figure can be compare with Fig.~3 (left panel) of~\cite{ppprd} where a similar trend arises because the spectra for all $\hat n_{ch}$ have been deliberately normalized to unit integral. The crossing feature in that case resulted from the interplay between soft and hard components with changing control parameter $\hat n_{ch}$.

As an initial approximation the spectra were renormalized with two-component factors $[1 + 0.08(\nu - 1)]$ for pions and $[1 + 0.16(\nu - 1)]$ for protons. The $x$ values are based on Phenix data~\cite{phenix}. After the two-component analysis was established the factors were adjusted to final numbers 1.07, 1.15, 1.23, 1.31 1.35 for pions and 1.05, 1.20, 1.42, 1.68, 1.80 for protons based on spectrum structure near $y_t \sim 2$, as indicated in the relevant figures.

\end{appendix}


\end{document}